\documentclass[usenatbib,useAMS]{mnras}
\usepackage{newtxtext,newtxmath}
\usepackage[T1]{fontenc}
\usepackage{ae,aecompl}
\usepackage{amsmath, amssymb}
\usepackage{algpseudocode}
\usepackage{algorithm}
\usepackage{graphicx}
\usepackage{subfigure}
\usepackage{lineno}
\usepackage{afterpage}
\usepackage{comment}
\usepackage{enumitem}
\usepackage{multirow}

\usepackage{psfrag}

\let\hat\widehat
\title[Detecting Galaxy-Filament Alignments]{
Detecting Galaxy-Filament Alignments in the Sloan Digital Sky Survey III
}
\author[Yen-Chi Chen et al.]{Yen-Chi Chen,$^{1}$\thanks{E-mail:
yenchic@uw.edu}
Shirley Ho,$^{2,3,4}$
Jonathan Blazek$^{5,6}$,
Siyu He,$^{2,3,4}$
Rachel Mandelbaum,$^{3,4}$\newauthor
Peter Melchior,$^{7}$
Sukhdeep Singh$^{8,2,3,4}$
\\
$^{1}$Department of Statistics, University of Washington, Seattle, WA 98195, USA\\
$^{2}$Lawrence Berkeley National Lab, Berkeley, CA 94720, USA\\
$^{3}$Department of Physics, Carnegie Mellon University, Pittsburgh, PA 15213, USA\\
$^{4}$McWilliams Center for Cosmology, Carnegie Mellon University, Pittsburgh, PA 15213, USA\\
$^{5}$Institute of Physics, Laboratory of Astrophysics, \'{E}cole Polytechnique F\'{e}d\'{e}rale de Lausanne (EPFL),1290 Versoix, Switzerland\\
$^{6}$Center for Cosmology and AstroParticle Physics, Department of Physics, Ohio State University, Columbus, OH 43210, USA\\
$^{7}$Department of Astrophysical Sciences, Princeton University, Princeton, NJ, 08544, USA\\
$^8$ Berkeley Center for Cosmological Physics and Department of Physics, University of California, Berkeley, CA 94720
}

\begin{document}

\pagerange{\pageref{firstpage}--\pageref{lastpage}} \pubyear{2017}

\maketitle

\label{firstpage}

\begin{abstract}
Previous studies have shown the filamentary structures in the cosmic web influence the alignments of
nearby galaxies. 
We study this effect in the LOWZ sample of the Sloan Digital Sky Survey
%We study the effect of filaments on the alignment of adjacent galaxies in the LOWZ sample of the Sloan Digital Sky Survey.
using the ``Cosmic Web Reconstruction'' filament catalogue.
We find that LOWZ 
galaxies exhibit a small but statistically significant alignment in the direction parallel to the
orientation of nearby filaments.  This effect is detectable even in the absence of nearby galaxy
clusters, which suggests it is an effect from the matter distribution in the filament.
A nonparametric regression model suggests
that the alignment effect with filaments extends over separations of $30-40$ Mpc. 
We find that galaxies that are bright and early-forming align more strongly with the directions of
nearby filaments than those that are faint and late-forming; however, trends with stellar mass are
less statistically significant, within the narrow range of stellar mass of this sample.
%We also observed a mildly significant effect from the stellar mass.
%In all these three properties,
%we observe a significant effect from these three properties. 

%the brightness -- brighter galaxies tend to align stronger
%along the orientation of nearby filaments.

%However, we do not find any significant effect from the stellar mass and age. 
%We also find that filaments do not have a noticeable effect on galaxy-cluster alignment.
%%To further analyze how galaxy's properties, such as brightness, stellar mass, and age, 
%%affect the galaxy-filament alignment.
%%we fit a double-exponential model and a nonparametric regression model.
%%Both models suggest that the effect extends over $30-45$ Mpc. 
%%Moreover, we analyze how galaxy properties, such as brightness, stellar mass, and age, 
%%affect the galaxy-filament alignment.
%%We observe, with $2-3 \sigma$ evidence, 
%%that bright and old galaxies tend to be aligned more strongly than faint and young galaxies, while variation in stellar mass does not significantly influence the alignment signal. 
%%We also find that filaments do not have a noticeable effect on galaxy-cluster alignment.
\end{abstract}

\begin{keywords}
%large-scale structure, filaments, galaxy clusters
(cosmology:) large-scale structure of Universe
\end{keywords}

\section{Introduction}

%\jab{look back at citations when document compiles correctly} 
Cosmological simulations of large-scale structure reveal a consistent picture for the alignments of
dark matter halo shapes and spins with filaments \citep[for a review of galaxy and halo alignments, see][]{2015SSRv..193....1J}.
The fact that filaments are associated with the large-scale tidal field
\citep{2007MNRAS.375..489H,2007MNRAS.381...41H, 2008MNRAS.383.1655S,2009MNRAS.396.1815F,
  2010MNRAS.408.2163A, Cautun2012}, which directly sources the dark matter halo spins through tidal
torquing and imprints coherent patterns in their shapes, makes this connection quite natural and
unsurprising.  For details of how dark matter halo spins and shapes align with filamentary
structures, and the dependence on other properties such as halo mass, see for example \citet{2006MNRAS.370.1422A,2007ApJ...655L...5A,hahn+07,2009ApJ...706..747Z,2013MNRAS.428.2489L,2014MNRAS.440L..46A,2018arXiv180500033G}.

However, the expectations for galaxy (not halo) shape and spin alignments are less obvious, and the
observational results are somewhat in conflict with each other.  For example,
\citet{2013ApJ...775L..42T} and \citet{2015ApJ...798...17Z} reported contradictory results for the
alignments of spiral galaxy spins with filaments.  The differing results may arise due to
differences in sample selection, spin measurement, or mass ranges.  %\rachel{Should perhaps collect
%  some more confusing observational results here.}\cite{2018arXiv180500033G}

In this work, rather than investigate galaxy {\em spin} alignments with filaments, we focus on
galaxy {\em shape} alignments with filaments.  The primary motivation behind considering shape alignments
is that coherent galaxy shape alignments with the cosmic web are a contaminant to weak gravitational
lensing measurements \citep[for recent reviews, see][]{2015RPPh...78h6901K,2017arXiv171003235M} that
serve as one of the most promising probes of dark energy.  Weak gravitational lensing involves
statistical measurements of low-level coherent galaxy shape distortions induced by the gravitational
potential of large-scale structure, and hence ``intrinsic alignments'' (coherent alignments due to
large-scale tidal fields) must be modeled and removed. There are several approaches to modeling
these alignments: to date, most cosmological weak lensing analyses
\citep[e.g.,][]{2017arXiv170801530D,2017MNRAS.465.1454H} have used numerical models for intrinsic
alignments that should be valid only to mildly nonlinear scales.  More sophisticated analytic
modeling schemes have been developed, some with additional nonlinear terms
\citep[e.g.,][]{2017arXiv170809247B}, which can include the impact of spin alignments, and others based on the inclusion of a halo model for the
effects that arise inside of dark matter halos \citep{2010MNRAS.402.2127S}.  It is important to
understand whether galaxy shapes are aligned with filaments in a way that requires inclusion of
additional terms in these models in order to properly remove intrinsic alignments contamination from
future weak lensing measurements.

In this paper, we study the effect from filaments on galaxy shape alignments using the LOWZ sample in the Sloan Digital Sky Survey (SDSS).
We acquire filaments from the cosmic web reconstruction catalogue \citep{2016MNRAS.461.3896C} and
use a catalogue of galaxy shapes measured using the
re-Gaussianization method \citep{Hirata2003}. 
%Section~\ref{sec::SDSS} briefly describes the construction of the filament catalogue and the measurement of galaxy shapes. 
Our results can be compared with a similar analysis using the same filament-finding method
in cosmological hydrodynamic simulations \citep{2015MNRAS.454.3341C}.  In addition to following
previous works in quantifying how galaxy-filament alignments scale with galaxy properties, we also
attempt to separate out the impact of filaments on galaxy alignments from the impact of clusters on
galaxy alignments.

The outline of the paper is as follows.  After describing the data used for this work in
\autoref{sec::SDSS}, we show the impact of filaments on 
on galaxy-cluster alignments in \autoref{sec::GC}. 
In \autoref{sec::GF}, we measure galaxy-filament alignments
after accounting for the alignment of galaxies towards clusters.
We study how galaxy-filament alignments scale with separation as well as the impact of different galaxy properties. 
%\autoref{sec::effect_radius} shows how galaxy-filament alignments scale with separation, while 
%\autoref{sec::property} is about the scaling with galaxy properties. 
Finally, we conclude in \autoref{sec::conc}.

	Throughout the paper, we assume a WMAP7 
$\Lambda$CDM cosmology with $H_0 = 70$~km/s/Mpc, $\Omega_m = 0.274$, 
and $\Omega_\Lambda = 0.726$
%\citep{2012MNRAS.427.3435A,2014MNRAS.439...83A} 
\citep{Komatsu2011}
%\rachel{I don't understand what is intended by these
%citations; it seems from context like you intend to cite the WMAP7 cosmology paper, but instead you
%are citing some BOSS analyses?  I recommend changing this.}  
%\sukhdeep{I have added WMAP7 paper. Yen-chi please check to make sure this is the one you are using and not some other WMAP paper.}
and use angular diameter distances for calculation
of physical distances.

\section{Data}\label{sec::SDSS}
	\subsection{The Sloan Digital Sky Survey}
		The SDSS 
		\citep{2000AJ....120.1579Y} 
		imaged roughly $\pi$ steradians
		of the sky, 
%and the SDSS-I and II surveys followed up approximately one million of the detected
%		objects spectroscopically \citep{2001AJ....122.2267E,
%		  2002AJ....123.2945R,2002AJ....124.1810S}. The 
		with the imaging carried
		out by drift-scanning the sky in photometric conditions
		\citep{2001AJ....122.2129H, 2004AN....325..583I}, in five bands
		($ugriz$) \citep{1996AJ....111.1748F, 2002AJ....123.2121S} using a
		specially-designed wide-field camera
		\citep{1998AJ....116.3040G} on the SDSS Telescope \citep{Gunn2006}. These imaging 
		data were used to create
		the  catalogues of galaxy shapes that we use in this paper. The SDSS-I/II imaging
		surveys were completed with a seventh data release
		\citep{2009ApJS..182..543A}, though this work will rely as well on an
		improved data reduction pipeline that was part of the eighth data
		release, from SDSS-III \citep{2011ApJS..193...29A}; and an improved
		photometric calibration \citep[`ubercalibration',][]{2008ApJ...674.1217P}.

		\subsection{Baryon Oscillation Spectroscopic Survey (BOSS)}
			Based on the photometric catalog from 
			SDSS,  galaxies are selected for spectroscopic observation 
			\citep{Dawson:2013}, and the BOSS spectroscopic survey was performed
			\citep{Ahn:2012} using the BOSS spectrographs \citep{Smee:2013}. Targets
			are assigned to tiles of diameter $3^\circ$ using an adaptive tiling
			algorithm \citep{Blanton:2003}, and the data were processed by an
			automated spectral classification, redshift determination, and parameter
			measurement pipeline \citep{Bolton:2012}.

			We use SDSS-III BOSS data release 12 \citep[DR12;][]{SDSS2015}
			LOWZ galaxies. This sample consists of 361,762 galaxies over an area of 8377 deg$^2$ \citep{2016MNRAS.455.1553R}.
			We use these galaxies to construct
			an overdensity map from which we construct the filament map. 

			To get the shapes of BOSS galaxies, we use the same shape catalog as was used in \cite{Singh2015}. The galaxy shapes were 
			measured by 
			\citet{Reyes2012} using the 
			re-Gaussianization method \citep{Hirata2003} of correcting for the effects of the point-spread function (PSF) on 
			the observed galaxy shapes. We refer the reader to \cite{Singh2015} and \cite{Reyes2012} 
			for further details of the shape measurements.
			
			We also use galaxy clusters from the redMaPPer catalogue version 10
	        \citep{2014ApJ...783...80R,2014ApJ...785..104R,2015MNRAS.450..592R}  to test the effects
            of clusters on the galaxy alignments. 
%            \rachel{Should say which version of the redMaPPer
%              catalogue was used, since there have been multiple iterations with different area
%              coverage and other properties.}
	        In addition we also use publicly-available estimates\footnote{\url{http://www.sdss.org/dr12/spectro/galaxy_granada/}} of galaxy stellar mass and 
	        age based on the Flexible Stellar Population Synthesis code of \cite{2009ApJ...699..486C}, in order to test the impact of these properties on galaxy alignments.

        \subsection{Filament Catalogue}	\label{sec::methods}
        We obtain filaments from
        the \emph{Cosmic Web Reconstruction}\footnote{\url{https://sites.google.com/site/yenchicr/}}
        catalogue 
        \citep{2016MNRAS.461.3896C},
        a publicly-available filament catalogue consisting of 
        filaments in SDSS from 
        redshift $z=0.05$ to $z=0.70$.
        Note that in this paper, we only use the LOWZ sample, so we restrict ourselves
        to $0.20<z<0.43$. 
        
%        Here we briefly describe the detection of filaments in 
%        the catalogue.
        The catalogue is constructed by
        first slicing the spectroscopic galaxy sample from redshift $z=0.05$ to $z=0.70$
        into $130$ thin slices along line of sight with width $\Delta z=0.005$
        and then projecting galaxies within each slice onto two dimensional plane of RA, DEC (denoted as $x$ in following discussion). 
        The galaxies in the same slice are then smoothed into a two dimensional probability density field, $p(x)$
        using a Gaussian kernel $K(x)$
%      \sukhdeep{Is there any reason why equations are not numbered in the paper?}
%      \peter{Yeah, these aren't equation environments, they are all done as display math.}
%      \sukhdeep{I understand. My question is if there is any specific reason (eg. some Journal specific requirements) that they are written this way?}
        \begin{equation}
        p(x) = \frac{1}{nh^2}\sum_{\ell=1}^n K\left(\frac{x-x^\ell}{h}\right)
	\end{equation}
        with the smoothing bandwidth $h>0$ chosen by the 
        reference rule described in \cite{2015MNRAS.454.1140C}.
        {{In the LOWZ sample, $h$ ranges from $20-40$ Mpc.}}
        Finally, we detect the filaments as the ridges of the density field \citep{2014arXiv1406.5663C}
        using the subspace constrained mean shift algorithm \citep{Ozertem2011}.
        
        { Note that the density field $p(x)$ is constructed assuming a flat space geometry 
        (i.e., we do not take into account the fact that the RA-DEC coordinate is curved) so there will be some 
        systematics caused by ignoring this geometry. However, such systematics are very tiny because our smoothing
        scale is about $1-2$ degrees. For a Gaussian kernel, contribution from $3\sigma$ away has
        a value less than $1/8000$ so they are negligible in the analysis. This suggests that the systematics from geometry
        is at the scale of comparing a short arc of $3-6$ degrees to a straight line connect the two ends of the arc. 
        This difference is $|\sin \left(\frac{6\degr}{2}\right) - \frac{6\degr}{2}|
        \approx\left(\frac{\pi}{60}\right)^3 \approx 10^{-3}=0.1\%$. 
        So the systematics from ignoring the geometry will not impact much on the density field $p(x)$.}
        
        The ridges are identified using the hessian of the density field ({tidal tensor}) ,
        $H_{ij}(x) = \frac{\partial^2}{\partial x_i\partial x_j} p(x)$
        with $v_1(x), v_2(x)$ as its eigenvectors corresponding to the eigenvalues $\lambda_1(x)\geq \lambda_2(x)$.
        The ridges are defined as
        \begin{equation}
        R = \left\{x: v_2(x)^T \nabla p(x) =0, \lambda_2<0\right\},
        \end{equation}
        where $\nabla p(x)$ is the gradient of the density field.
        $R$ is the collection of points where the gradient projected onto the subspace spanned by the second eigenvector $v_2(x)$
        is zero and the second eigenvalue is negative. 
        This implies that every point on $R$ is a local maximum in the subspace spanned by $v_2(x)$,
        so $R$ can be viewed as a curve consisting of many local maxima in some subspaces, which is the feature of a ridgeline. 
%        More details on the construction algorithm can be found in 
%        \cite{2016MNRAS.461.3896C}.

        Note that ridges (hereafter filaments) are 1-dimensional objects (curves),
        so we can easily define the orientation of every point on a filament. 
        In reality, a filament is represented by a collection of points
        and every point
        contains a vector $\eta_{\sf filament}$ indicating
        the orientation of the filament passing through the point.
        {
        Points on filaments are created from applying the ridge finding algorithm
        (the subspace constrained mean shift algorithm) to a uniform 2D grid. 
        Roughly speaking, the ridge finding algorithm moves points on the 2D grid according to the 
        gradient of galaxy's distribution until these points arrive on ridges 
        (please see Figure 2 of \citealt{2015MNRAS.454.1140C} for an illustration). 
        Note that we approximate the orientation of filaments using the density gradient so
            it may be slightly different from the actual orientation; the details of the approximation is in \cite{2016MNRAS.461.3896C}. 
        The average separation between points on a filament is about 0.77 Mpc.
        This can be viewed as the uncertainty due to the resolution of our filament finder.
        This uncertainty is much smaller than the uncertainty of filament's location due to sampling 
        (around 10-20 Mpc; see
        \citealt{2016MNRAS.461.3896C}) so we can ignore its effect on the alignment signal.
        Note that increasing the grid density will improve the resolution of filament finder but will also increase 
        the computational cost. We choose the grid density to be dense enough but still computationally feasible. }

        \begin{figure}
	        \includegraphics[height=2in]{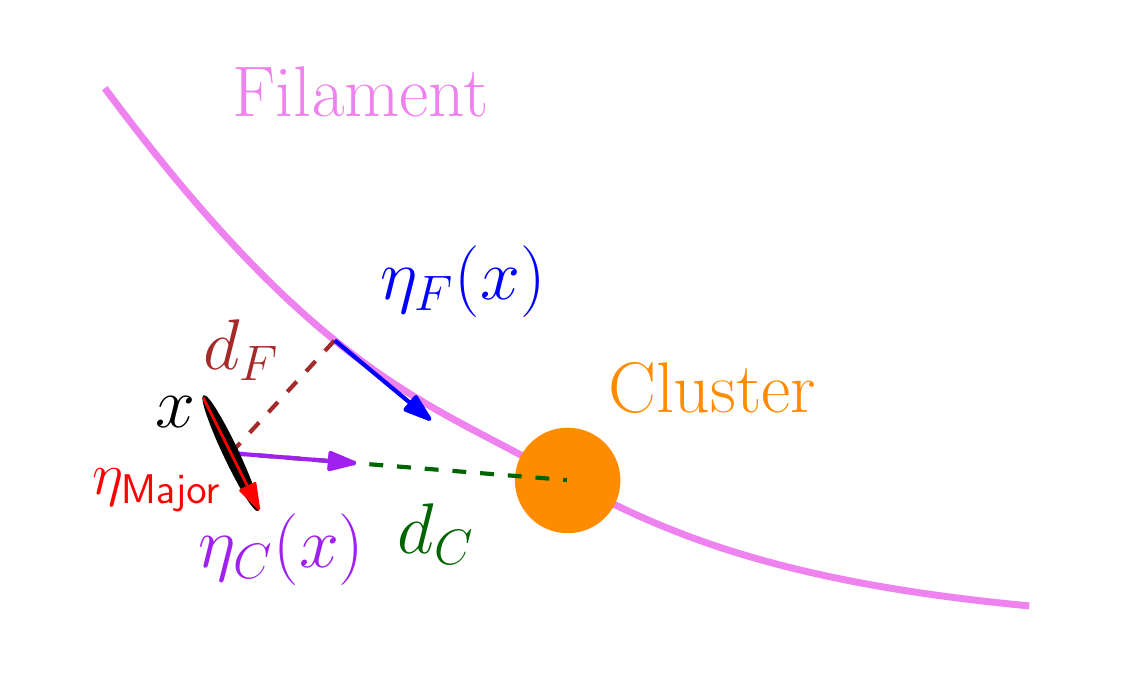}
    	    \caption{For a galaxy located at position $x$, this figure illustrates the quantities
              $\eta_F$ (orientation of the nearby filament at the
              point nearest to the galaxy) and $\eta_C$ (orientation of the unit vector connecting
              the galaxy and a cluster).
	        The distance to the filament ($d_F$) is the length of the brown dashed line
    	    and the distance to the cluster ($d_C$) is the length of the green dashed line.
            Finally, $\eta_{\sf Major}$ shows the major axis direction of the projected galaxy shape.}
        	\label{eq::ex01}
        \end{figure}
        
        \begin{figure}
	        \includegraphics[height=3.3in]{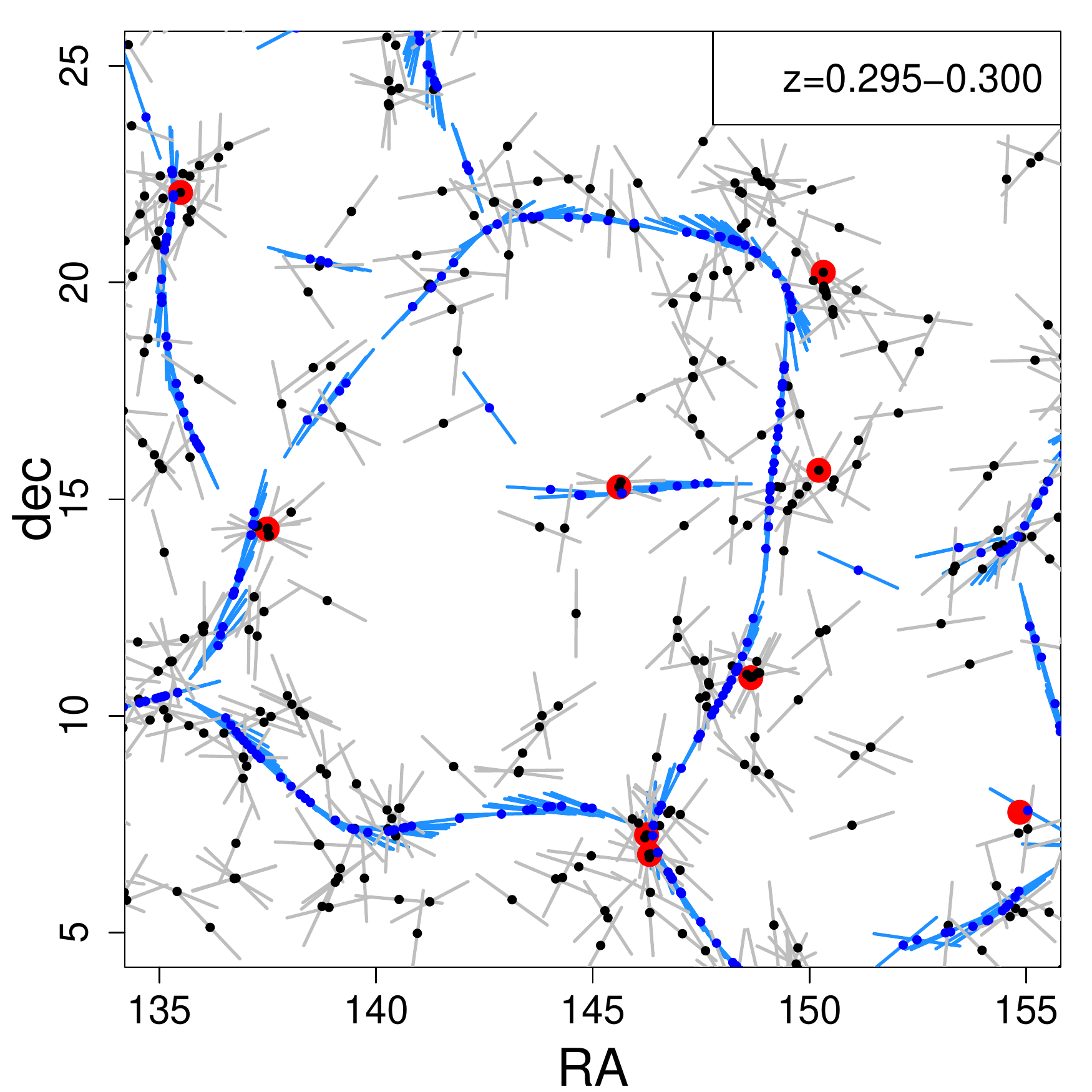}
    	    \caption{
        	An illustration of galaxies (black dots), clusters (red dots), and filaments (curves, represented by dark blue dots)
	        in a narrow redshift slice in the LOWZ sample. 
%\rachel{This is pretty confusing, because the text says that
%              filaments are {\em curves}.  You say that in reality it is represented by a set of
%              points, but I think that people reading the caption without reading that one sentence
%              of the text will find this statement that filaments are represented by dots
%              confusing.}
%\sukhdeep{added a small text}
    	    The gray lines on the galaxies indicate the direction of their major axes.
        	The light-blue line segments indicate the orientation of the filament at the location of
            the dark blue dots. 
            One degree on this figure corresponds to $\sim 20$~Mpc given the redshift indicated on
            the legend.
	        }
    	    \label{eq::ex02}
        \end{figure}
        
        For each galaxy, we denote $d_F$ as its distance to
        the nearest point representing filaments. This quantity will be a proxy for
        the distance to the closest filament. One an view 
        $d_F$ as the projected distance from a galaxy to the nearest filament in terms of angular
        diameter distance. 
        More specifically, filaments in the Cosmic Web Reconstruction catalogue are constructed
        in each redshift slice. For a given galaxy, we first identify which
        redshift slice that it belongs to. 
        Then we compute its distance to the nearest filament in the same slice, disregarding the
        fact that galaxies near the borders between slices may be physically closer to filaments in
        an adjacent slice.  This simplification may slightly reduce the significance of the observed alignment
        effects. 
%\rachel{Should say how the nearest filament was determined when factoring
%        in 3D distance (i.e., did you calculate a 3D distance including redshift information, or 2D within narrow redshift
%        slices, or \dots?)}
% \sukhdeep{I believe this is 2-D; projected distance. But Yen-chi should clarify this.}
%\rachel{So there could be a more nearby filament along the line-of-sight?}
        Given a galaxy located at position $x$, % (RA, DEC),
        we denote $\Pi_F(x)$ as its projected point onto the filament.
        Then the vector
        \begin{equation}
        \eta_F(x) = \eta_{\sf filament} (\Pi_F(x))
	\end{equation}
        is defined as the orientation of the filament (unit vector tangential to the filament) nearest to the galaxy located at $x$.
        Similarly, for each galaxy, we define $d_C$ and $\eta_C$ as the distance and unit vector to the nearest redMaPPer cluster.
%Thus, $\eta_C$ is the orientation toward the nearest cluster.\sukhdeep{this statement is not correct. Orientation is defined using 
%        shape of galaxies.}
        \autoref{eq::ex01} provides an illustration of the quantities $\eta_F$, $\eta_C$, $d_F$, and $d_C$ for
        a given galaxy, filament, and cluster.
        For each galaxy, the primary shape-related quantity we are interested in is 
        $\eta_{\sf Major}$, the direction of the major principal axis of the galaxy projected onto the plane of the sky.
%        \jab{(projected onto the plane of the sky)}. 
%        \sukhdeep{$\eta_{\sf Major}$ is not shown in the figure}
        
        In \autoref{eq::ex02} we show the distribution of galaxies, filaments and their shape orientations in a small 
        region within the LOWZ sample ($0.295<z<0.300$). Our focus will be on understanding the alignments of the galaxies (gray lines)
        and the influence of filaments (blue lines) and clusters (red points) on these alignments. 
        A notable feature in the figure is that the galaxies are primarily distributed close to the 
        filaments and clusters, with clusters themselves being close to the filaments. This becomes important as we later split the galaxies into 
        subsamples based on distances to filaments and clusters, where the samples with larger distance from clusters and filaments will have fewer 
        galaxies, which increases the noise in the measurements.
%   \sukhdeep{This is incomplete. What do we want the reader to take away from this figure?}
%   \sukhdeep{Update: I have added my main takeaway from this figure. If there is something more, please do add it here.}
%        The galaxies are shown by black dots, clusters by red dots, and filaments by blue dots. 
%        The major axis of each galaxy is given by the gray line segment
%        and the orientations of filaments are shown as the light-blue line segments.\sukhdeep{this is in caption. No need to repeat the thing}

%Our analysis tries to study how the gray segments and light-blue segments are aligned. 

%For each galaxy, we consider two shape-related quantities:
%$\eta_{\sf Major}$ and $r_{ab}$.
%$\eta_{\sf Major}$ is a 2D unit vector indicating the direction of the
%major principal axis of this galaxy and $r_{ab}$ is the axes ratio
%of the minor principal axis and the major principal axis.

\section{Results}
\subsection{Filament effect on galaxy-cluster alignment}	\label{sec::GC}

Several studies have been performed to understand the galaxy alignments with surrounding density field using both observations and simulations 
%\citep{2006ApJ...644L..25A,2007ApJ...662L..71F, 1989ApJ...344..535W,2009ApJ...703..951W,
%2011MNRAS.414.2029P,2012MNRAS.423..856S, 2015SSRv..193..139K}. 
\citep[see eg. ][ for review]{2015SSRv..193..139K}
%\rachel{Someone needs to fix these
%references; they are clearly not the right ones for this sentence.  For example, West et al and Smargon et al are
%about cluster-cluster alignments, not galaxy-cluster alignments; Kirk et al is a review article
%about all aspects of galaxy intrinsic alignments, rather than a study of galaxy-cluster alignments.
%(It's OK to cite a review article, but don't imply that it's a study of galaxy-cluster alignments!)}
%\sukhdeep{I have removed most the references and also changed the language a bit.}
In this section, we investigate the effects of filaments on the galaxy-density alignment using clusters as the tracers of density field. We quantify the galaxy-cluster alignment with the statistic \citep{2009ApJ...706..747Z,2013MNRAS.428.1827T, 2015MNRAS.450.2727T,2015MNRAS.454.3341C}:
{
\begin{equation}\label{eq:cl-alignment-stat}
|\phi_C-\phi_{\sf Major}| ={\sf arccos}(|\eta_C\cdot \eta_{\sf Major}|).
\end{equation}
Namely, the quantity $\phi_C-\phi_{\sf Major}$ is the angular difference between the the two
axes $\eta_C$ and $\eta_{\sf Makor}$.
It can be computed by inverting
the inner product $\eta_C\cdot \eta_{\sf Major}$.  
Note that since we are comparing the angular difference of two axes,
the value occurs between $0$ degree (perfectly aligned) to $90$ degrees (perfectly anti-aligned). 
%measures the alignment between the direction
%of the major axis of the galaxy, $\eta_{\sf Major}$ and the direction towards the cluster
%$\eta_{C}$. 
%Since for galaxy alignments the positive and negative value of inner product implies the
%same degree of alignment (rotating an ellipse by 180 degrees leads to the same ellipse), we measure
%the absolute values. 
For the case of random alignments, this quantity will be (uniformly) randomly distributed 
over the interval $[0,90]$.
%this inner product averages to $2/\pi$.
}
We note that for the galaxies with low ellipticity the estimates of the direction of the major axis 
become noisier (for a round galaxy, the major axis direction is ill-defined). However, we do not apply any 
preferential treatment to low ellipticity galaxies in our analysis as this effect mainly adds to noise to 
our measurement.
%The advantage of using this cosine statistic is that the alignment is easy to interpret -- 
%it is just the amplitude of the inner product between two vectors.
%The cosine statistic was also used in other literature
%for measuring the alignment between two angles; see, e.g.,
%\cite{2009ApJ...706..747Z,2013MNRAS.428.1827T, 2015MNRAS.450.2727T,2015MNRAS.454.3341C}.

\begin{figure*}
\includegraphics[height=1.8in]{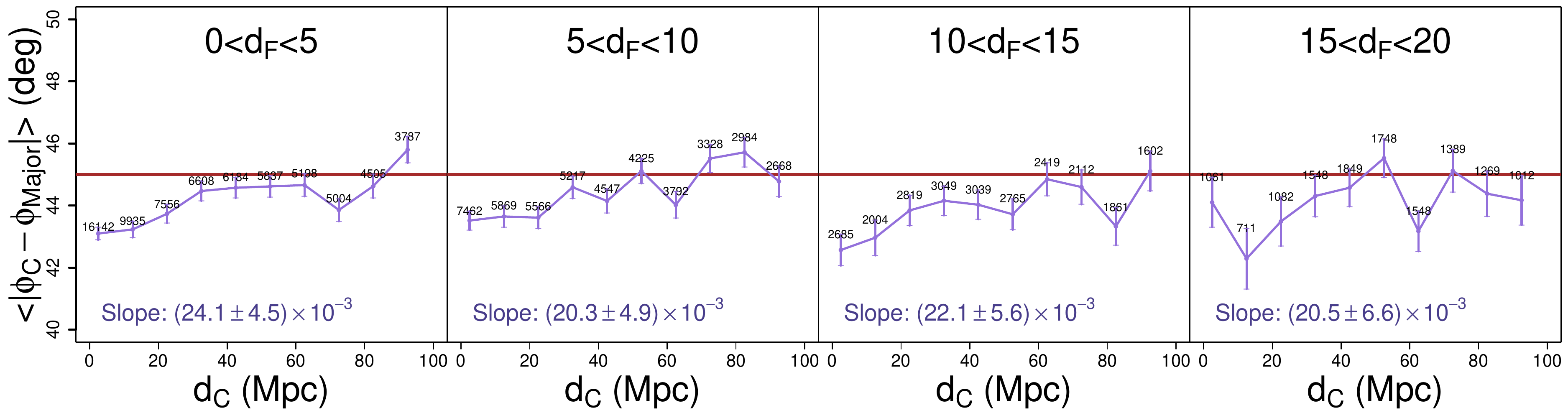}
\includegraphics[height=1.8in]{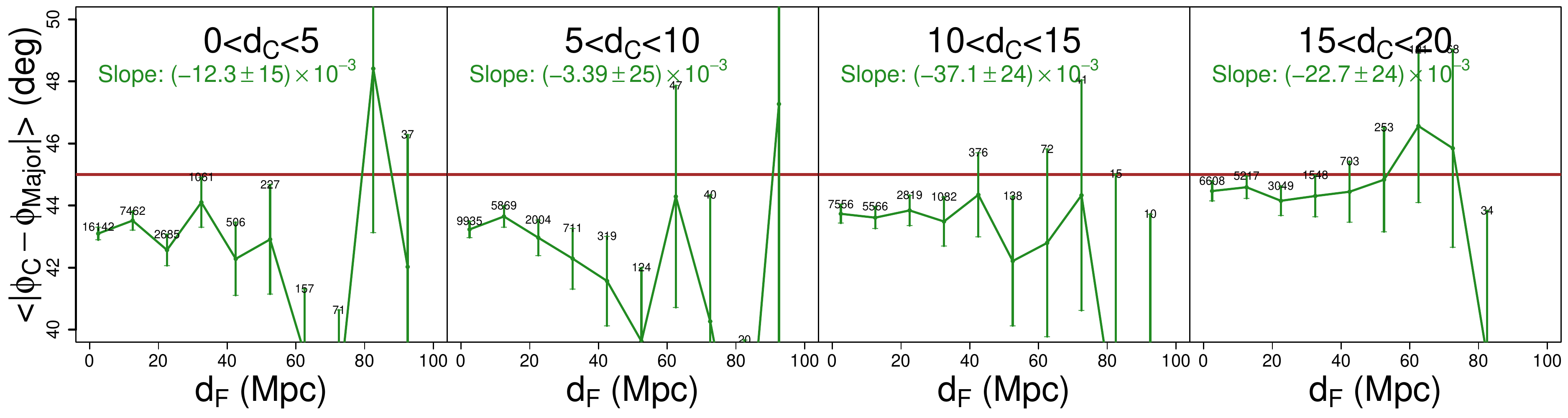}
\caption{
The galaxy alignment signal toward clusters $\langle|\phi_C-\phi_{\sf Major}|\rangle=\langle{\sf arccos}|\eta_C\cdot \eta_{\sf Major}|\rangle$ 
as a function of distance to the clusters, $d_C$ and filaments, $d_F$. 
{ $\langle \cdot \rangle$ denotes the mean of the statistic. }
%\sukhdeep{There is small notational inconsistency here. Inside the panels distance are denoted by symbols, $d_c,d_F$ and the axis label it is in full text. Perhaps we should be more consistent?}
%\sukhdeep{Also, it will be good to mentions units of distances, preferably on figures themselves or atleast in captions.}
The numbers on the bins denotes the number of galaxies in each bin.
The brown horizontal line indicates the average value from a random alignment. 
{\bf Top panel:} the alignment towards clusters as a function of projected distance to the cluster at
various (fixed) distances to filament as indicated on the plot.
The slope is from fitting a linear regression to the scale- (distance-) dependence of the signal.
In the first two panels, there is significant scale dependence of the alignment signal toward clusters. In the next two panels, measurements are noisier as the number of galaxies in bins decreases, making it harder to infer the scale dependence of the signal.
{But in all cases, we see that the curve is going up, indicating that clusters do have impact 
on the alignment of nearby galaxies.}
{\bf Bottom panel:} the alignment toward clusters as a function of distance to filament
at various (fixed) distances to the clusters.
In this case we do not observe a significant scale dependence. Results from both panels suggest that
the filaments do not have a significant impact on the alignment of galaxies toward clusters, within the uncertainty in the measurements.
{The fact that most parts of the curves are below 45 degrees (i.e., an alignment effect)
is because we are only looking at galaxies that are very close to clusters (they are all below $20$ Mpc to the cluster).
These galaxies are greatly influenced by the effect of cluster. 
}
%\sukhdeep{Sorry, I think I asked this before, but why only focus on the slopes? Given that the noise in the bins is order $10^{-2}$, I don't know how meaningful are the slope estimates of $10^{-5}$.}
%\rachel{The vertical axis label is wrong (the $\langle\rangle$ are omitted).  Also, need to say what
%the horizontal line is.}
}
\label{fig::align_dFdC_C2}
\end{figure*}

We study the {average} alignment $\langle|\phi_C-\phi_{\sf Major}|\rangle$ 
in two cases. (i): we fix $d_F$ and study how $\langle|\phi_C-\phi_{\sf Major}|\rangle$ changes as $d_C$ changes; 
(ii): we fix $d_C$ and study how $\langle|\phi_C-\phi_{\sf Major}|\rangle$ changes as $d_F$ changes.
\autoref{fig::align_dFdC_C2} shows the results for both cases;
the error bars are based on the standard error of the average within each bin.
In all cases, a statistically significant alignment of galaxies towards clusters is seen, but with
different scale dependence.  
%\jab{About this notation: $\mu$ is commonly used as the $\cos$ of an angle, often the angle between a separation direction and the line of sight. There may be some (mild) confusion due to this convention.}
%%YC: I changed it to \eta; hopefully this is better
In the top row, we display the result of case (i).
%The slope is based on a fitted linear regression the observations.
In the two leftmost panels ($0$ Mpc $<d_F<10$ Mpc), we observe a significant dependence of the
alignments on $d_C$, 
%suggesting the existence effect from distance to cluster $d_C$ on the galaxy-cluster alignment.
with 
the strength of the galaxy alignment toward the cluster decreasing as the distance to the nearest cluster increases.  
%This is affirmative to the previous literatures that galaxies around clusters tend to align toward their nearby clusters.
%This scale dependence of the measurements is 
This decreasing pattern 
is qualitatively consistent with the galaxy alignments literature where the alignments between galaxies and the density field follows a power law relation
%The relation between the alignment strength and the distance approximately follows a power law, 
%{qualitatively} consistent with the other observations of intrinsic alignments in the literature
\citep[see e.g. ][ for a review of galaxy alignment measurements]{2015SSRv..193..139K}. 
The significance of the galaxy-cluster alignment measurements 
decreases as the distance to the filaments increases (two rightmost panels in top row of \autoref{fig::align_dFdC_C2}).
This trend is primarily driven by the increased noise as the number of galaxies decreases with the increased distance to filament and 
does not imply a significant filament effect on the galaxy-cluster alignment.

In the bottom row, we show the result of case (ii) to study the galaxy-cluster alignment as function
of the distance of the galaxies to the filaments.
We do not observe any significant dependence of galaxy alignments towards galaxy clusters on the
distance to the filaments. 
These results suggest that filaments do not have a significant impact on the tendency of galaxies to
align towards galaxy clusters.
%, suggesting that such effects have to be smaller than the noise in our measurements. 

%\rachel{Need to say where the errorbars on the plots come from.}
%% YC: I added when we first mentioned \autoref{fig::align_dFdC_C2}.

\subsection{Galaxy-filament alignment}
\label{sec::GF}
%\jab{we could probably combine sections 4-6. A paper of this length doesn't need 7 sections.}

\begin{figure}
\center
\includegraphics[height=2.4in]{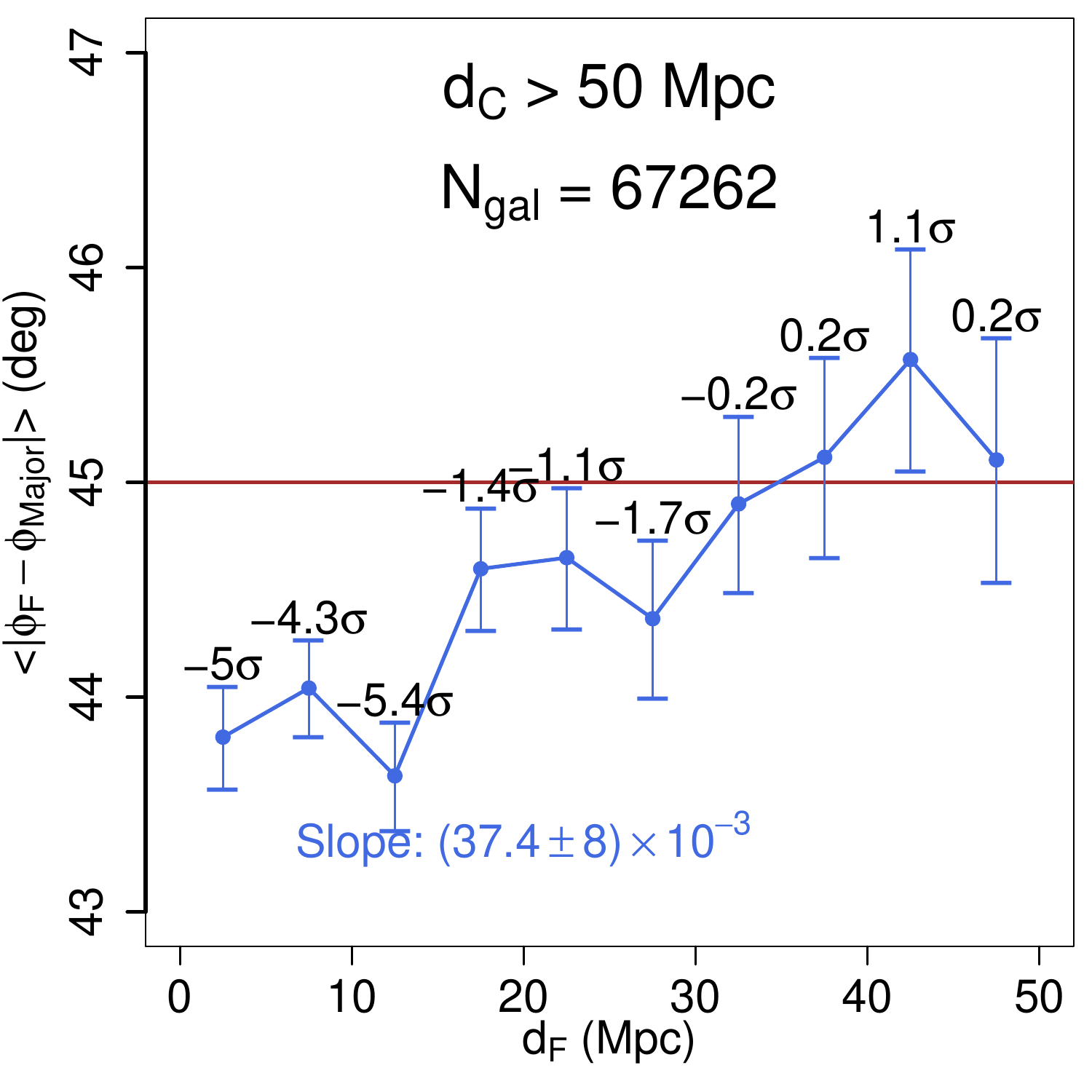}
\caption{The galaxy-filament parallel alignment signal $\langle|\phi_F-\phi_{\sf Major}|\rangle$
as a function of distance to the filament for galaxies with a cluster-centric distance $d_C$ above 50\,Mpc. 
%10\,Mpc (left), 30\,Mpc (center), 50\,Mpc (right). 
We observe statistically significant galaxy-filament parallel alignment that
decreases with separation from the filament, with a significant non-zero slope as indicated on the plot. 
The brown horizontal line indicates the average value from a random alignment. 
%\rachel{The vertical axis label is wrong (the $\langle\rangle$ are omitted).  Also, need to say what
%the horizontal line is.}
}
\label{fig::align}
\end{figure}
%\sukhdeep{Need to have consistent notation for distances in all figures}

% Note: sentence below talked about alignment "towards" filaments.  That's not what we're measuring;
% we measure alignments in the parallel direction to the filament direction, not alignments in the
% sense of "galaxies point towards the direction of the nearby filament".  I tried to change
% language consistently.
In this section we investigate whether galaxies are aligned in parallel to nearby filaments. We define the filament orientation as the direction of the tangent at minimum separation from the galaxy (see \autoref{eq::ex01}), which gives 
{
the alignment statistic
\begin{equation}
\langle|\phi_F-\phi_{\sf Major}|\rangle = {\sf arccos}(|\phi_F-\phi_{\sf Major}|).
\end{equation}}
%\rachel{I think this needs an absolute value or a square inside of the $\langle\rangle$, similar to \autoref{eq:cl-alignment-stat}?}
Similar to the galaxy-cluster alignment measurement above,
we now study how the galaxy-filament alignment changes as a function of 
distance to filaments $d_F$.

\subsubsection{Separating impact of filaments and clusters}
As shown in \autoref{sec::GC}, clusters influence the alignments of galaxies up to the maximum scale
measured in this work, $d_C<50$ Mpc. Figure~\ref{fig::align_dFdC_C2} shows 
that the alignment effect of galaxies towards clusters is consistent with random at this distance 
%\jab{should it be 10 or 50 here?}
%% YC: i will just use 50 to be more conservative -- we do see that 10 Mpc might already be enough
%(the distance that alignment signal drops to the average magnitude of random alignments),
%\jab{where does this number come from?}
and we therefore consider only galaxies with larger distances from the nearest cluster. 
\autoref{fig::align} shows the alignment of galaxies in the parallel direction with respect to
nearby filaments given a 
cluster-centric distance 
threshold of $>50$ Mpc.
%three cluster-centric distance 
%thresholds: 
%$>10$ Mpc, $>30$ Mpc, and $>50$ Mpc.  
%\rachel{This sentence completely contradicts the previous one.  You just said you only want to
%  consider galaxies more than 50 Mpc from the cluster, but now you say that \autoref{fig::align}
%  considers separations down to 10 Mpc!}  
 The parallel alignment signal is measured as a function of
$d_F$, the distance to the nearby filament.
In every panel, we see statistically significant detection at small values of $d_F$, with a
decreasing alignment up to $30-40$ Mpc from the filament; 
linear regressions over the entire range in $d_F$ yield significant detection of a negative slope in all three cases.
The cut with $d_C>50$ Mpc is particularly relevant because at this distance, the galaxy--cluster alignment becomes negligible (see \autoref{fig::align_dFdC_C2}).
Our interpretation of this finding is that filaments induce a coherent alignment signal that is not directly associated with clusters.

\subsubsection{Range of galaxy-filament alignment effect}
\label{sec::effect_radius}

\begin{figure}
\center
\includegraphics[height=6in]{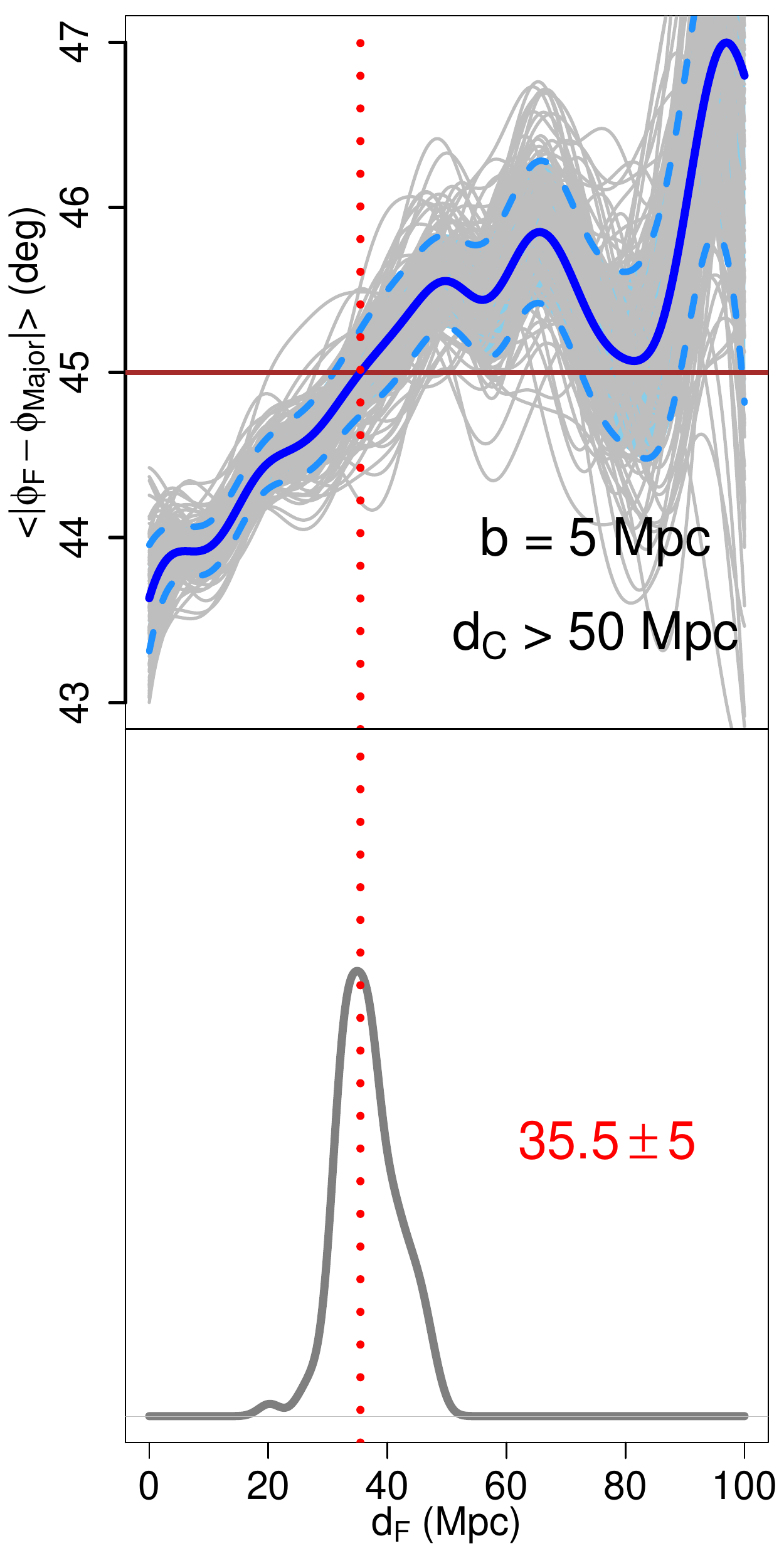}
\caption{The range $\rho_0$ of the galaxy-filament parallel alignment effect for galaxies with a
  cluster-centric distance $d_C$ above 50\,Mpc. % (right). 
%10\,Mpc (left), 30\,Mpc (center), 50\,Mpc (right).
{\bf Top:} The local polynomial regression fits with smoothing bandwidth $b=5$ Mpc for $100$ bootstrap samples (gray); their mean (blue); and their 90\% pointwise confidence bands (cyan).
%\rachel{The vertical axis label is wrong (the $\langle\rangle$ are omitted).}
{\bf Bottom:} The probability density plot of $\rho_0$ from $100$ bootstrap samples.
The vertical dotted red line indicates the distance that the alignment signal first dropped to the average of random alignments.
%\rachel{Vertical axis label should say ``Probability density of $\rho_0$''.  Also, say what vertical
%dotted line is.}
}
\label{fig::NP1}
\end{figure}

%\sukhdeep{We should perhaps note somewhere here that $\rho_0$ could be due to uncertainty in filaments and does not necessarily imply some physical scale.}
To determine the range of the filament alignment effect $\rho_0$,
%We now want to determine the range of the filament alignment effect $\rho_0$
%, and define the \emph{effective radius of filaments} $\rho_0$ 
%\jab{is ``effective radius'' confusing, since we're not talking about the radius of the filament itself?} 
%% \YC{I changed it to the range of the filament alignment effect. }
we use
the cutoff distance where the 
angular difference goes over $45$ degrees (average of random alignment). 
%drops below the value of random alignment $2/\pi\approx 0.637$. 
%Note that $2/\pi$ is a heuristic choice that does not have a physical meaning because we do not have a model for the phenomenon. 
Also note that  the value of $\rho_0$ could be due to uncertainty in filaments and does not necessarily imply some physical scale

%\jab{Why is this the relevant threshold? Unlike with clustering, there is no physical reason why it would need to drop below random. So where it ``crosses'' could just be due to noise. The sampling method accounts for noise, but do we have an explanation for what is going on? Put another way: anti-alignment is not the same as ``no effect.''}

We use the local polynomial regression \citep{wasserman2006all} as a non-parametric method to estimate the regression function.
Given a value of covariate $x$ (here the covariate is the distance to filament $d_F$), 
the local polynomial regression weights all the data points based on the distance from their covariate values to $x$. 
Close data points are given higher weights and far away data points are given less weights.
In the local polynomial regression, there is a smoothing bandwidth $b$ that determines how the weight as a distance to $x$ decays.
Roughly speaking, points within $[x-b, x+b]$ will have a much higher weight compared to those outside this window. 
One can view this approach as a modified method of fitting a regression based on taking the average over a sliding window of width $2b$.
The reasons of using a local polynomial regression are that (i) it has a smaller statistical error compared to the conventional binning approach, and (ii) it does not require any prior knowledge about the shape of the regression function. 

We present the results in \autoref{fig::NP1} with a bandwidth $b = 5$ Mpc; an analysis with different bandwidths yields comparable results.
%From the left to the right columns we consider different thresholds on $d_C$.
%\rachel{Based on the previous subsubsection, it seems that only the rightmost panel is relevant?}
The top panel displays the fit results for $100$ bootstrap samples; the bottom panel shows the distribution of $\rho_0$ based on those bootstrap samples.
In agreement with \autoref{fig::align}, we find the range $\rho_0\approx 35$ Mpc with a standard error of about $5$ Mpc. 

\begin{figure}
\center
\includegraphics[height=2.8in]{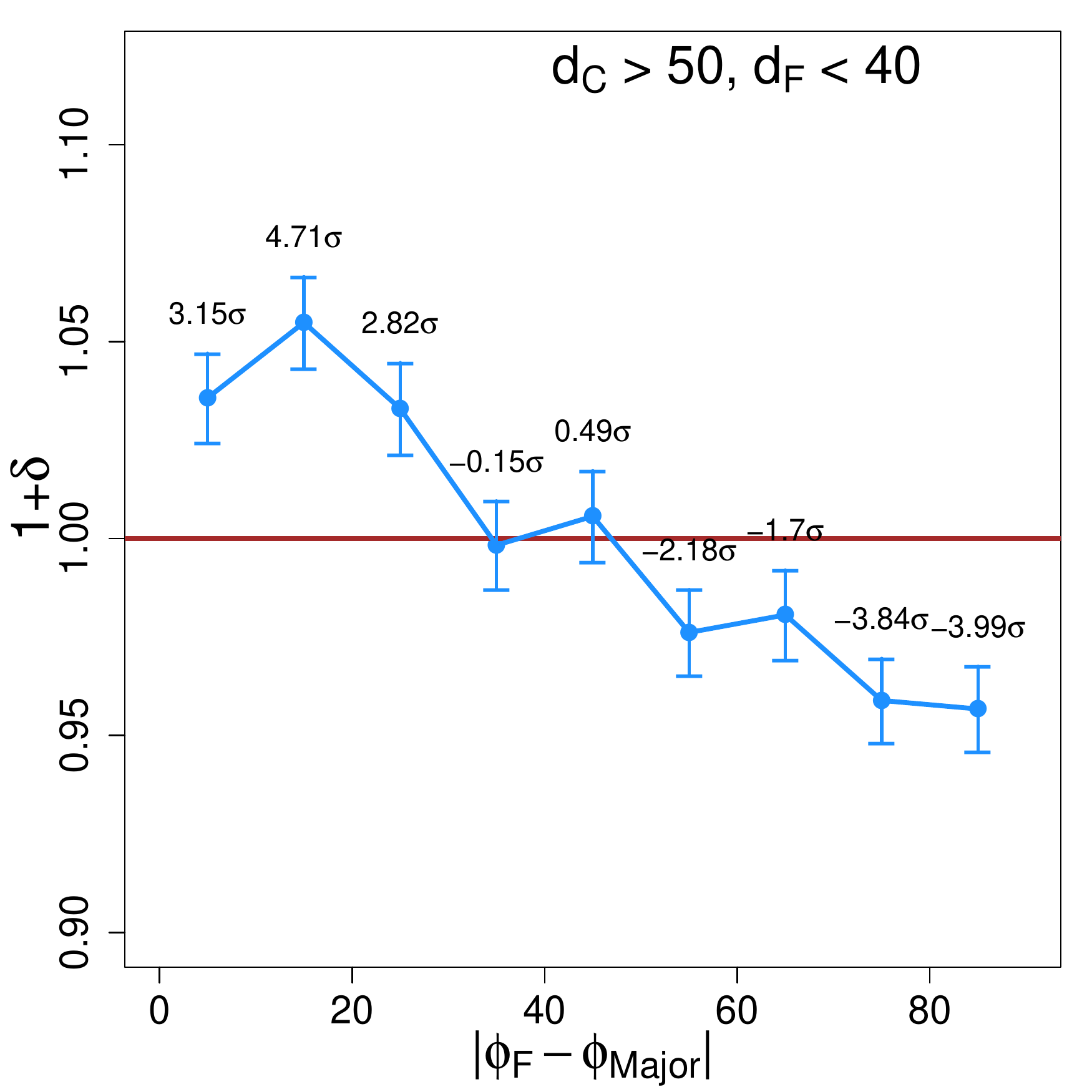}
\caption{
Excess probability density of the alignment in the parallel direction to nearby filaments for galaxies within the range of the filament alignment effect $\rho_0$ 
and cluster-centric distances $d_C$ above 50\,Mpc. 
$\delta = p_{\sf obs}/p_{\sf rand}-1$, where $p_{\sf obs}$ is the probability density 
from observed alignments and $p_{\sf rand}$ is the probability density from random alignments.
%10\,Mpc (left), 30\,Mpc (center), 50\,Mpc
%(right). 
%\rachel{Vertical axis label has quantity $\delta$ that has not been defined.  I think the
%  horizontal axis label is missing an absolute value?}
%\peter{I strongly suggest that we show only one of those rows.}
}
\label{fig::align_cosine}
\end{figure}

Using the range of the filament alignment effect, we calculate the excess probability density
function of the difference angle  $|\phi_F-\phi_{\sf Major}|$, 
%\rachel{Is this missing an absolute value?}
normalized to a pair of random axes.
%\peter{This text could use an equation}.
%\YC{Because we use the cosine statistics, the formula I know could be a bit ugly so I choose not to use it.}
We only use galaxies within 40\,Mpc, i.e. those that are within the range of the effect, with
cluster-centric distances $d_C>50$ Mpc.
%$d_C>10$, $30$, and $50$ Mpc. \rachel{Again, based on previous sections, I
%have concerns about the two left panels.  You also never really comment on the way the results
%depend on $d_C$, which suggests to me that we don't need all the panels.} 
The results are shown in \autoref{fig::align_cosine}, demonstrating that there are more galaxies aligned along their nearby filaments than one would expect from uniform random orientations.

%%%%% This is what was deleted due to referee's request
%A similar result has been observed in \cite{2015MNRAS.450.2727T} and \cite{2013ApJ...779..160Z}.
%The excess probability rises to about $1.05$, which is similar to the findings of \cite{2013ApJ...779..160Z} but somewhat smaller than the result of \cite{2015MNRAS.450.2727T}. 
%In \autoref{sec::property}, we will make a more detailed comparison to these two works. 
%\jab{What does this excess probability reveal that the average angle statistic does not?}
%% YC: we are just trying to compare to what Zhang et al 2013 and Tempel 2015 had reported; in their paper they use the excess probability plot. 

{{Note that one may notice the anti-alignment at large $d_F$ (blue curve went over
the brown horizontal line) in the top panel of Figure~\ref{fig::NP1}. 
Such an anti-alignment may be caused by the fact that when a galaxy is far away from a filament, 
the nearest filament can be viewed as a point source to this galaxy. 
We expect a galaxy to align toward a point mass, which explains why we observe the anti-alignment.
%To examine if this comes from systematics in our analysis,
To further examine such an effect, 
we analyze the alignment of those galaxies with $d_F>50, 100$ Mpc 
in \autoref{sec::anti}. 
When we investigate those galaxies, we do not observe a significant alignment.
So we cannot conclude if such an anti-alignment is a realistic effect from our study.
%xxx
%xxx
%xxx
}

%which suggests that the anti-alignment is probably due to randomness of the sample.}

\subsubsection{Galaxy properties and galaxy-filament alignment}
\label{sec::property}

\begin{figure*}
\includegraphics[width=6in]{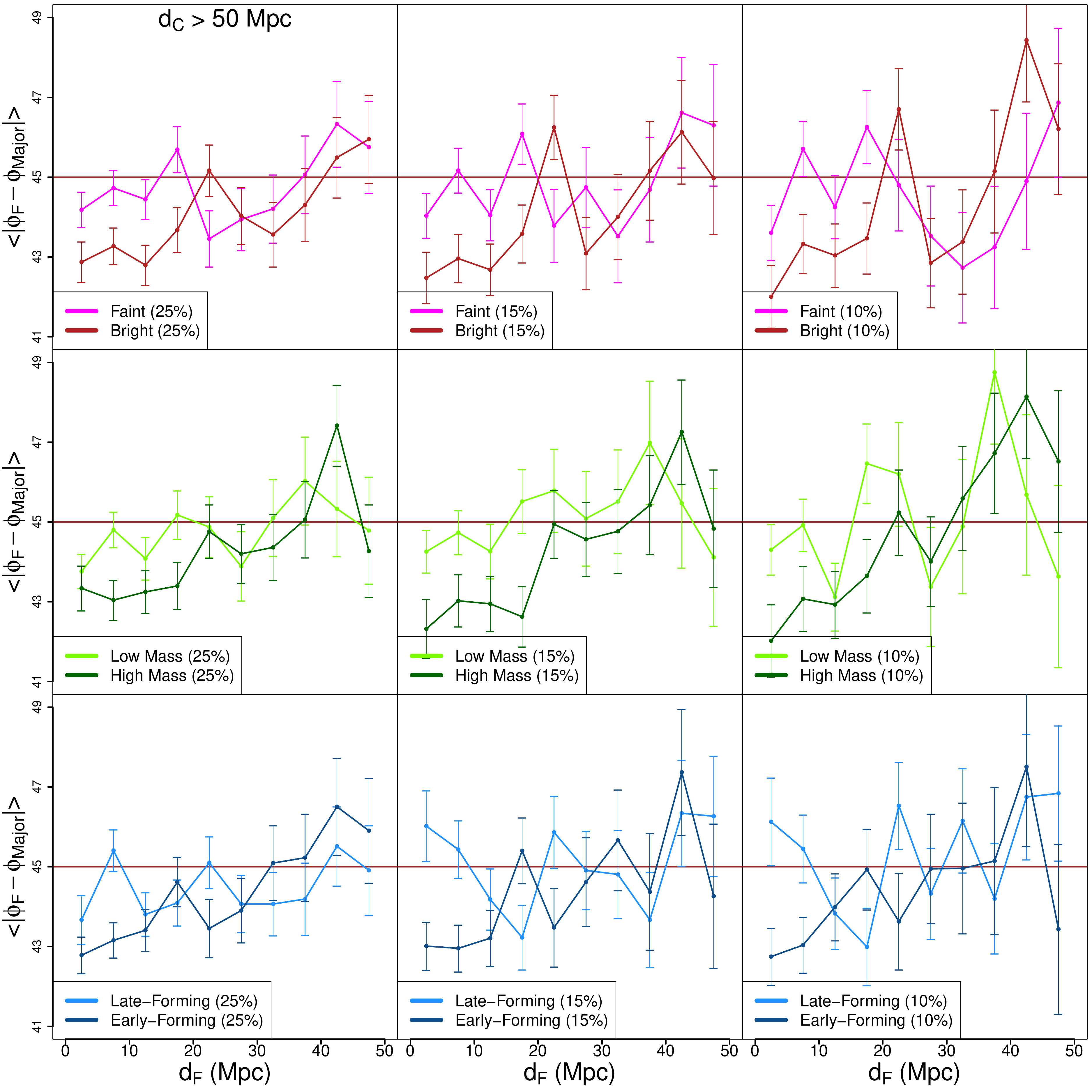}
\caption{This figure shows the effect of galaxy properties on filament alignment. 
We compare the two most extreme groups of galaxies according to a particular galaxy property.
From left to right, the extreme groups are selected with $25\%$, $15\%$, and $10\%$ criteria. In all
panels, we only consider galaxies with $d_C > 50$ Mpc to reduce the impact of cluster alignments.
The p-value for the difference between the two extreme groups is given in \autoref{tab::test1}.
{\bf Top row:} We separate galaxies by their brightness ($r$-band absolute magnitude).
The upper and lower brightness thresholds are $(-22.16, -21.69)$, $(-22.31,-21.58)$, and $(-22.42,-21.50)$. 
%%% avg: -21.45, -22.44 // -21.38, -22.57 // -21.30, -22.68
{\bf Middle row:} We separate galaxies by their stellar mass.
The upper and lower mass thresholds are $\text{log}_{10}(M_*/M_\odot)=(11.76, 11.55)$, $(11.83,11.50)$, 
and $(11.87,11.47)$. 
%%% avg:: 10.25, 10.96 // 10.03, 11.03 // 9.90, 10.08
{\bf Bottom row:} We separate galaxies by their age.
The upper and lower age thresholds are $(8.38, 7.06)$, $(8.73, 6.59)$, and $(9.13, 6.41)$ Gyr.
%\jab{since all panels use 30 Mpc, that number can just be displayed once.}
%%% avg:: 6.98, 9.62 // 6.46, 9.89 // 6.13, 10.10
}
\label{fig::property}
\end{figure*}

%\clearpage
%this is a temporary hack to deal with the latex float issue

\begin{figure*}
\includegraphics[height=2.4in]{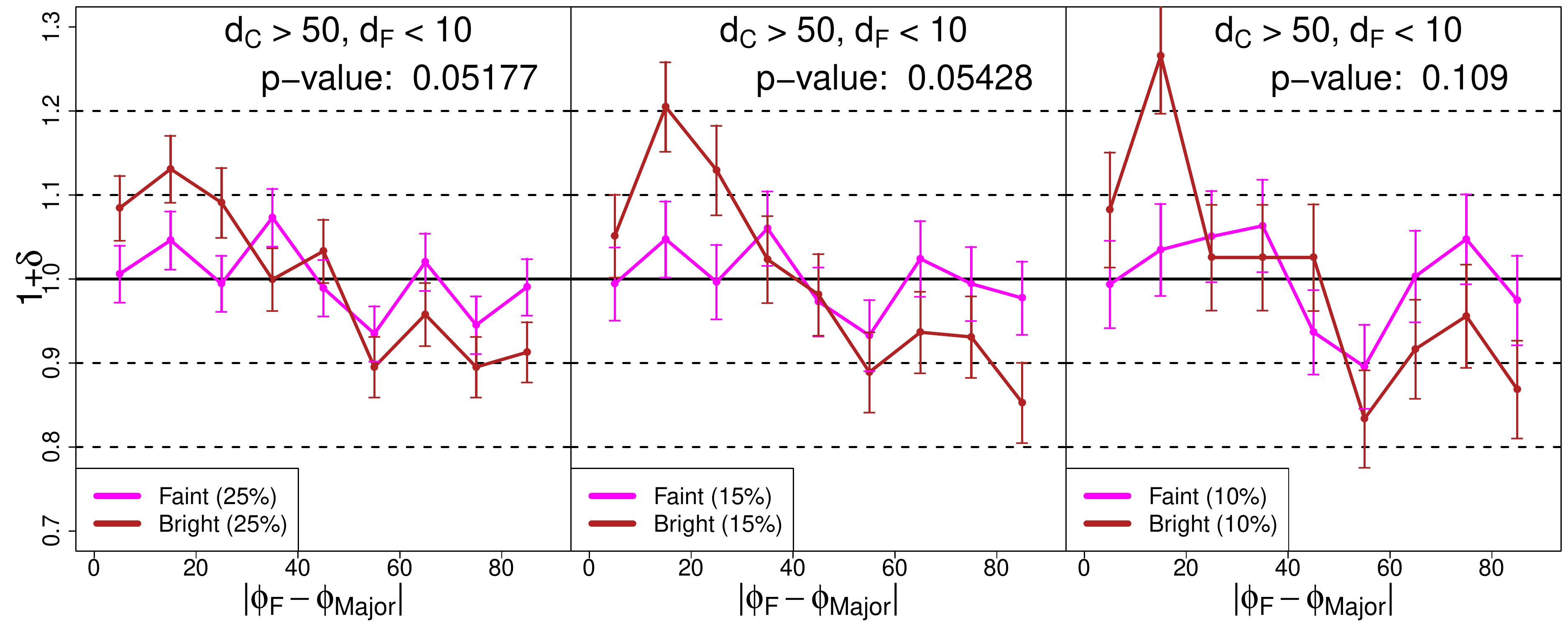}\\
\includegraphics[height=2.4in]{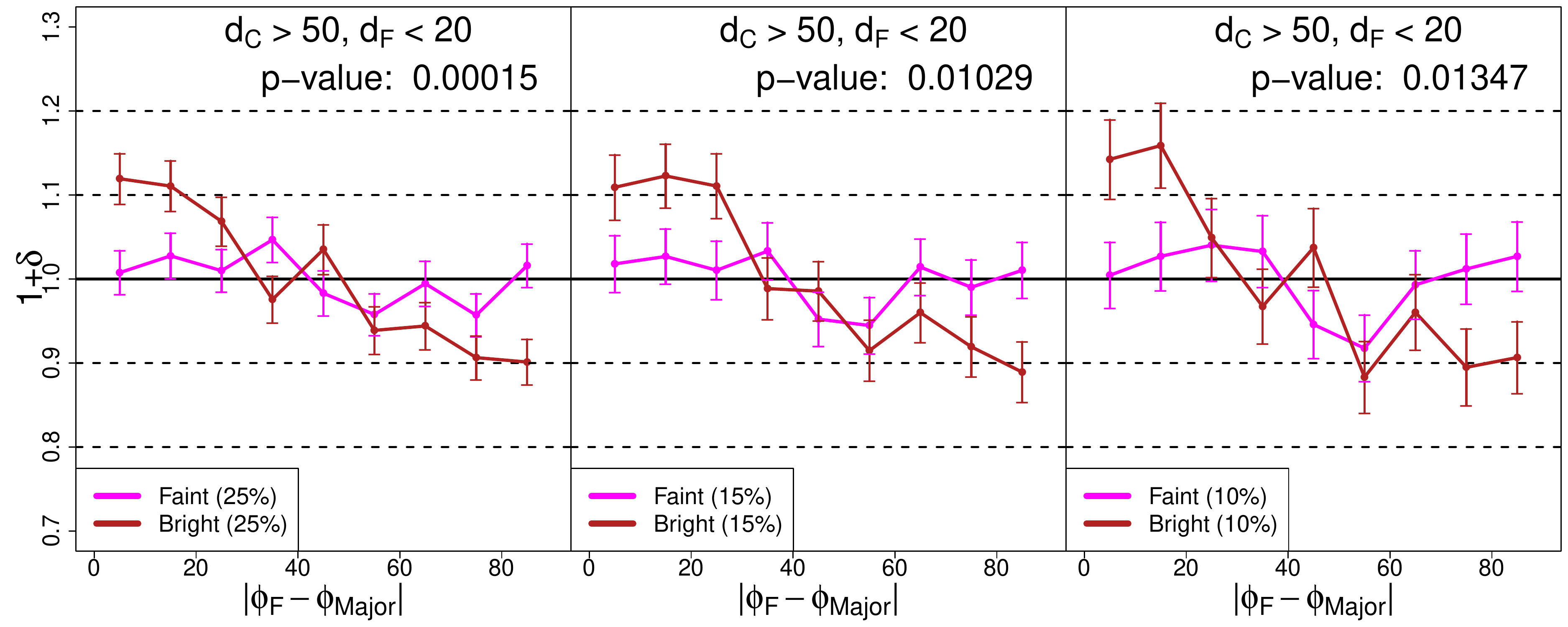}
\caption{This figure shows the effect of brightness on the excess probability of galaxy-filament alignment. 
This figure is similar to the bottom row of \autoref{fig::align_cosine}, but here we compare the
brightest and dimmest galaxies while thresholding the maximum distance to filaments at $10$ Mpc ({\bf top row}) and $20 $ Mpc ({\bf bottom row}).
%\jab{[why the difference in threshold?]}.
%\YC{[just to show that we have tried different settings and the results remain similar. Maybe we can just go with the top row and say that
%other threshold gives a similar result?]}
Note that the top right panel shows the p-value computed from a 
two-sample $\chi^2$ test
to examine whether the two groups (brightest versus faintest) have the same
excess probability. 
%where $\chi_9^2$ indicates that these are for 9 degrees of freedom. 
%\rachel{Similar to my previous comments, what is the value of showing $\chi_9^2$ versus p-values?
%  If you only show $\chi^2$ then people have to go look up the p-values themselves.}
%\jab{??}
%% \YC: I added more explanation here.
We compare the brightest galaxies versus the faintest galaxies using the $r$-band magnitude. 
We observe a significance increase in the excess probability at smaller misalignment angles, indicating the presence of
galaxy-filament alignment.
Moreover, when we consider more extremely separated galaxy populations, the increase in excess
probability at small misalignment angle seems to increase in amplitude (top row, from the left panel to the right panel). 
%\jab{is this a significant effect? the chi2 goes down for the more extreme binning.}
%% \YC: It was significant for the second bin in the top row (the error bar does not intersect in the first and the third panel) 
%% but not an overall patten so I just rephrase this claim.
}
\label{fig::align_cosine_lum}
\end{figure*}

\begin{table}
\centering
\begin{tabular}{crrr}
  \hline
 & 25\% & 15\% & 10\% \\ 
  \hline
%Brightness & 27.54 & 21.89  & 14.93  \\ 
%Stellar Mass & 8.7  & 14.88  & 9.65  \\ 
%  Age & 16.42  & 25.81  & 13.83  \\ 
Brightness & $\mathbf{7.1\times 10^{-4}}$ & $\mathbf{4.8\times 10^{-3}}$  & $\mathbf{2.9\times 10^{-2}}$  \\ 
Stellar Mass & $7.5\times 10^{-2}$  & $\mathbf{1.1\times10^{-2}}$  & $5.0\times10^{-2}$  \\ 
  Age & $\mathbf{3.0\times10^{-2}}$  & $\mathbf{3.3\times10^{-3}}$  & ${6.8\times10^{-2}}$  \\ 
   \hline
\end{tabular}
\caption{P-value of testing the difference between the two extreme groups in Figure~\ref{fig::property}
using a $\chi^2$ test.
We highlight the case where the p-value is less than $0.05$, a conventional threshold
for claiming significance. 
%\rachel{I am a bit confused; the age splits seem the least convincing, yet they are more significant
%by far than the mass splits.  Do you understand what is driving this?  Is it just the single point
%that is closest to the filament?}
%The degree of freedom of the $\chi^2$ statistics is $8$.
}
\label{tab::test1}

\end{table}

To investigate whether the galaxy-filament alignment depends on galaxy properties, 
we perform the same analysis after separating galaxies 
%\jab{[just the shape sample, right? the filament catalog is always the same?]} 
%%\YC: yes we do not change the filament catalog.
by their absolute magnitude,
stellar mass, and age. 
% \jab{size or age? It said ``size'' before, but that property is not mentioned again.}
%% \YC: it should be age, not size. 
The stellar mass and age are obtained from the Granada group catalog
\citep{2014ApJS..211...17A}.\footnote{
More details can be found at \url{http://www.sdss.org/dr13/spectro/galaxy_granada/}}
%RM: I changed the reference.  Conroy+ is just a reference for the templates they used, but not the
%reference for the stellar mass estimates.  They are described in the DR10 paper, Ahn+, which is now
%cited here.
%\peter{I think you should cite that relevant paper here.}

When binning on each of these three properties in turn, we compare the two most extreme groups for a
given property. 
For the brightness ($r$-band absolute magnitude), we compare the $25\%/15\%/10\%$ brightest galaxies versus the $25\%/15\%/10\%$ faintest galaxies, and 
likewise for stellar masses and ages.
Note that the distribution of these properties in the LOWZ sample is relatively narrow, which means
that even the extreme split performed here does not result in a very large difference between the samples.

%\sukhdeep{I believe this is still with $d_C>50$ cut? This cut could preferentially be throwing away brighter/older/more massive galaxies. This would further narrow the range of values of these properties. You can probably test this by comparing the distributions of these properties for full sample vs the one with $d_C$ cut.}
%\YC{The thresholds are constructed based on the LOWZ sample before the $d_C$ cut. So changing $d_C$ will not affect the thresholds.
%I checked the result by requiring the thresholds after $d_C$ cut and the results are almost the same ($p$-value changed so little that \autoref{tab::test1} does not change)}
%\sukhdeep{Thank you for checking. Intuitively it seems a bit weird. I just checked my older results and there also we had observed that central galaxies in the groups tend to be brighter but this effect is almost perfectly cancelled by the satellites which tend to be fainter. As a result the average luminosity for group galaxies was very similar to full sample (for LOWZ). I guess something similar might be at play here.}

%\autoref{fig::property} shows the alignment signals for the two extreme groups when we separate galaxies
%by each property.
% RM: you more naturally introduce this figure later, so let's leave it for there.

To quantify the significance of the different alignment behavior between the two extreme groups of galaxies,
we perform a simple $\chi^2$ test by binning galaxies according to their
distance to filaments ($d_F$).
{{In what follows we describe the details of the $\chi^2$ test we are using but essentially, 
it is the conventional $\chi^2$ test with the bins described in \autoref{fig::property}.}
Let $B_1,\cdots,B_8$ be the first $8$ bins of $d_F$ in
\autoref{fig::property}. 
Namely, $B_\ell$ contains galaxies with $5(\ell-1)\leq d_F<5\ell$ Mpc, where $\ell$ is the bin
index, ranging from 1--8.
These bins corresponds to the galaxies with $d_F<40$ Mpc. 
The upper bound $40$ Mpc which is roughly the range of filament's effect $\rho_0$; see \autoref{sec::effect_radius}. 
We use the following $\chi^2$ statistic to test if the two extreme groups have consistent alignment behavior:
\begin{equation}
\chi^2 = \sum_{\ell=1}^8 \frac{\left(\langle|\phi_F-\phi_{\sf Major}|\rangle_{\sf high,\ell}-\langle|\phi_F-\phi_{\sf Major}|\rangle_{\sf low,\ell}\right)^2}{\sigma^2_{\sf high,\ell}+\sigma^2_{\sf low,\ell}},
\end{equation}
where $\langle|\phi_F-\phi_{\sf Major}|\rangle_{\sf high,\ell}$ is the average alignment
of the highest group galaxies in bin $\ell$ and $\langle|\phi_F-\phi_{\sf Major}|\rangle_{\sf low,\ell}$
is that of the lowest group in bin $\ell$.
If there is no effect from the specified galaxy property, 
we expect no statistically significant difference between the two extreme groups,
and the
%If the two extreme groups are from the same population,
$\chi^2$ statistic will follow a $\chi^2_8$ distribution ($\chi^2$ distribution with degree of freedom $8$).
%We quantify the significance using $\chi^2/8$; namely, if a signal is $x\cdot \sigma$, the corresponding $\chi^2$ statistic is $\chi^2 = 8x$.
The results of this test for different galaxy properties and split criteria are displayed in \autoref{tab::test1}.

In \autoref{fig::property}, we show on the impact of galaxy properties on alignment with filaments.
In the top row, we split galaxies on absolute magnitude and observe a separation between the two groups of galaxies, with bright galaxies more aligned along nearby filaments compared to faint galaxies. The $\chi^2$ test shown in \autoref{tab::test1} confirms this pattern -- the two samples have a difference with p-values ranging
from $7.1\times10^{-4}$ to $2.9\times 10^{-2}$.
%$1.87-3.44\sigma$ ($\chi^2_8 = 14.93-27.54$) separations, depending on which criteria is used to classify the extreme groups. The corresponding p-values range from $6\%$ ($1.87\sigma$) to $0.06\%$ ($3.44\sigma$).

To further analyze how the brightness affects the alignment, 
we consider galaxies within 10 or 20 Mpc of a filament %(40 Mpc is the approximated 
%effect radius of filaments; see Section~\ref{sec::effect_radius})
and at least 50 Mpc away from a cluster.
We show the excess probability in terms of the angular difference 
$|\phi_F-\phi_{\sf Major}|$ in \autoref{fig::align_cosine_lum}.
%Note that the alignment signal is just the cosine of the angular difference. 
%If the alignment is random, then the angular difference will be uniformly distributed over $0\degr-90\degr$.
For both filament distance thresholds shown in \autoref{fig::align_cosine_lum}, we observe a clear increase in excess probability at small angular difference, implying that the major axes of both dim and bright galaxies tend to align along the orientation of nearby filaments. 
Moreover, the bright galaxies have a higher excess probability
at the small angular difference, indicating that 
the galaxy-filament alignment is stronger for bright galaxies.

The age of a galaxy also has a significant impact on the alignment (bottom row of \autoref{fig::property}).
Early-forming galaxies that are close to a filament tend to align
along the orientation of that filament whereas
the alignment of 
late-forming galaxies seems to be independent of filaments.
The $\chi^2$
%-square \sukhdeep{Need to use consistent notation for $\chi^2$} 
test in \autoref{tab::test1} shows a difference with p-values
ranging from $3.3\times 10^{-3}$ to $6.8\times 10^{-2}$.
However, this effect may just be due to the effect of brightness and the correlation between age and brightness.
Because we have a limited sample size, we do not adjust for the effect from brightness.
When we separate galaxies by their age, the resulting galaxies have an age-brightness correlation ranging from $0.24$ to $0.37$. 
Thus, when partitioning on age of galaxies, the early-forming galaxies tend to be brighter than the late-forming galaxies
and this brightness difference may cause the age effect. 

%$3.23-1.73\sigma$,
%which corresponds to a p-value of $0.11-8.64\%$.

%Filaments have a strong influence over early-forming galaxies -- galaxies close to filaments 
%Early-forming galaxies have 

The effect from stellar mass is more involved.
%On one case (when we compare the $15\%$ extreme groups), 
%we see a significant result (middle )
The middle row of \autoref{fig::property} shows some small separations between
the low mass and high mass galaxies in the sense that massive galaxies seem to have a stronger alignment 
effect.
But the $\chi^2$ analysis in \autoref{tab::test1} leads to a result that is close to the boundary
of significance.
The $\chi^2_8$ statistics give p-values ranging from $1.1\times10^{-2}$ to $7.5\times 10^{-2}$,
which is around the conventional significance level $0.05$. 
Hence we detect some possible effects from the stellar mass, but the results are not as significant
as those for brightness or age. 

{{Note that in \autoref{sec::other}, we apply the KS test and AD test (Anderson-Darling test)
to compare the two most extreme groups. 
The results remain very similar to what was observed in \autoref{tab::test1}. }

\section{Conclusion}	\label{sec::conc}

In this work, we have studied how filamentary structure impacts the orientation of galaxies in the
SDSS LOWZ sample.  We quantified two different types of galaxy alignments:
galaxy-cluster alignments and galaxy-filament alignments. 
Our primary results are as follows:
%\jab{removed the ``positive'' and ``negative'' terminology, which is less common in astrophysics.}
\begin{itemize}
\item {\bf We find a statistically significant galaxy-filament alignment.}
In the analysis of \autoref{sec::GF}, we observed a significant tendency for galaxies to be oriented
in a direction parallel to the nearest filament. 
%from filament on galaxy-filament alignment using the LOWZ sample.
Such an effect exists when we consider galaxies at least $50$ Mpc away from clusters, to remove the
impact of galaxy alignments towards clusters. 
Moreover, the analysis in \autoref{sec::effect_radius} further indicates
that the range of filament influence on the alignment is $\sim30-45$ Mpc, though some of this effect
appears to be determined by uncertainty in filament positions rather than a real alignment over tens
of Mpc. 
%Galaxies will tend to align along the orientation of filaments within this range. 

\item {\bf Galaxy brightness (absolute magnitude) and age impact the amount of alignment with filaments.}
Our analysis in \autoref{sec::property}
shows that the brightness and age of a galaxy impact the effect of filaments on alignment, with
brighter galaxies and/or those that are early-forming tending to be more aligned, especially when
the distance to the filament is $<10$ Mpc.
%more along the orientation of nearby filaments compared to dimmer galaxies. 
Overall, our results are qualitatively consistent with previous work (\citealt{2015MNRAS.450.2727T}
and \citealt{2013ApJ...779..160Z}) { although the scale is very different}. 
%In some scenarios, we observe a stronger effect than reported in
%those studies, although this may be due to differences in the galaxy samples and filament detection methodology.
%Note that the brightness effect may be due to the fact that 
%dim galaxies are difficult to measure their alignments.

\item {\bf We do not observe a significant impact of filaments on galaxy-cluster alignments.}
In the analysis of \autoref{sec::GC}, we find that the alignment of galaxies around clusters can be well explained by
the distance to the cluster itself, and there is no statistically significant contribution from filaments. 
However, we note that such an effect may exist at a level too weak to be detected in our data.
%imply the effect does not exist. 
%It is still possible that filaments do affect galaxy alignment
%but this effect might be too weak to be detected based on our data.

\item {\bf We find weak evidence for the galaxy-filament alignment strength to depend on stellar mass.}
When we separate galaxies according to their stellar mass, 
we only observe a difference in the galaxy-filament alignment that is on
the boundary of significance. 
We do not observe a very significant effect
as the 
%
%we do not observe a statistically significant difference in the galaxy-filament alignment (\autoref{sec::property}).  
%This result appears to be inconsistent with the 
findings of \cite{2009ApJ...706..747Z} and \cite{2015MNRAS.454.3341C}.
However, as before, differences in galaxy samples and methodology may account for this difference. 
%In particular, the difference of galaxy types (the LOWZ sample are mostly massive, red galaxies),
%small galaxy number density resulting in lower signal to noise, and different mass ranges
%are all possible causes.

Our results indicate that galaxy alignments are impacted by the complexity of large-scale structure beyond the positions of the most massive objects (clusters). We leave for future work an assessment of whether existing models for galaxy alignment, which typically treat correlations with the total density field or within the halo environment, are sufficient to account for the impact of filaments. Models that explicitly treat density-filament correlations may improve the overall description of galaxy alignments. Potential inconsistencies with previous results highlight the value of performing similar analyses on future data sets with greater statistical power.
%\jab{what do we want to say here in terms of future work/plans, if anything?}

\end{itemize}

%This paper analyzed the effect of filament on galaxy's alignment. 
%In particular, 

\section*{Acknowledgments}
We are grateful for the useful comments made by the referee. 
YC is supported by NIH grant number U01 AG016976. 
SH is supported by NASA ROSES grant 12-EUCLID12-0004 and NASA ROSES grant 15-WFIRST15-0008 and NSF-AST 1517593. 
JB acknowledges the support of an Ambizione Fellowship from the Swiss National Science Foundation.
RM is supported by NSF grant number AST-1716131.

Funding for SDSS-III has been provided by the Alfred P. Sloan Foundation, the Participating Institutions, the National Science Foundation, and the U.S. Department of Energy Office of Science. The SDSS-III web site is http://www.sdss3.org/.

SDSS-III is managed by the Astrophysical Research Consortium for the Participating Institutions of the SDSS-III Collaboration including the University of Arizona, the Brazilian Participation Group, Brookhaven National Laboratory, Carnegie Mellon University, University of Florida, the French Participation Group, the German Participation Group, Harvard University, the Instituto de Astrofisica de Canarias, the Michigan State/Notre Dame/JINA Participation Group, Johns Hopkins University, Lawrence Berkeley National Laboratory, Max Planck Institute for Astrophysics, Max Planck Institute for Extraterrestrial Physics, New Mexico State University, New York University, Ohio State University, Pennsylvania State University, University of Portsmouth, Princeton University, the Spanish Participation Group, University of Tokyo, University of Utah, Vanderbilt University, University of Virginia, University of Washington, and Yale University.

\bibliographystyle{mnras}
%\bibliography{SuRF_analysis.bib}
\bibliography{SDSS_FG}

\begin{thebibliography}{}
\makeatletter
\relax
\def\mn@urlcharsother{\let\do\@makeother \do\$\do\&\do\#\do\^\do\_\do\%\do\~}
\def\mn@doi{\begingroup\mn@urlcharsother \@ifnextchar [ {\mn@doi@}
  {\mn@doi@[]}}
\def\mn@doi@[#1]#2{\def\@tempa{#1}\ifx\@tempa\@empty \href
  {http://dx.doi.org/#2} {doi:#2}\else \href {http://dx.doi.org/#2} {#1}\fi
  \endgroup}
\def\mn@eprint#1#2{\mn@eprint@#1:#2::\@nil}
\def\mn@eprint@arXiv#1{\href {http://arxiv.org/abs/#1} {{\tt arXiv:#1}}}
\def\mn@eprint@dblp#1{\href {http://dblp.uni-trier.de/rec/bibtex/#1.xml}
  {dblp:#1}}
\def\mn@eprint@#1:#2:#3:#4\@nil{\def\@tempa {#1}\def\@tempb {#2}\def\@tempc
  {#3}\ifx \@tempc \@empty \let \@tempc \@tempb \let \@tempb \@tempa \fi \ifx
  \@tempb \@empty \def\@tempb {arXiv}\fi \@ifundefined
  {mn@eprint@\@tempb}{\@tempb:\@tempc}{\expandafter \expandafter \csname
  mn@eprint@\@tempb\endcsname \expandafter{\@tempc}}}

\bibitem[\protect\citeauthoryear{{Abazajian} et~al.,}{{Abazajian}
  et~al.}{2009}]{2009ApJS..182..543A}
{Abazajian} K.~N.,  et~al., 2009, \mn@doi [\apjs]
  {10.1088/0067-0049/182/2/543}, \href
  {http://adsabs.harvard.edu/abs/2009ApJS..182..543A} {182, 543}

\bibitem[\protect\citeauthoryear{{Ahn} et~al.,}{{Ahn} et~al.}{2012}]{Ahn:2012}
{Ahn} C.~P.,  et~al., 2012, \mn@doi [\apjs] {10.1088/0067-0049/203/2/21}, \href
  {http://adsabs.harvard.edu/abs/2012ApJS..203...21A} {203, 21}

\bibitem[\protect\citeauthoryear{{Ahn} et~al.,}{{Ahn}
  et~al.}{2014}]{2014ApJS..211...17A}
{Ahn} C.~P.,  et~al., 2014, \mn@doi [\apjs] {10.1088/0067-0049/211/2/17}, \href
  {http://adsabs.harvard.edu/abs/2014ApJS..211...17A} {211, 17}

\bibitem[\protect\citeauthoryear{{Aihara} et~al.,}{{Aihara}
  et~al.}{2011}]{2011ApJS..193...29A}
{Aihara} H.,  et~al., 2011, \mn@doi [\apjs] {10.1088/0067-0049/193/2/29}, \href
  {http://adsabs.harvard.edu/abs/2011ApJS..193...29A} {193, 29}

\bibitem[\protect\citeauthoryear{{Alam} et~al.,}{{Alam}
  et~al.}{2015}]{SDSS2015}
{Alam} S.,  et~al., 2015, \mn@doi [\apjs] {10.1088/0067-0049/219/1/12}, \href
  {http://adsabs.harvard.edu/abs/2015ApJS..219...12A} {219, 12}

\bibitem[\protect\citeauthoryear{{Altay}, {Colberg}  \& {Croft}}{{Altay}
  et~al.}{2006}]{2006MNRAS.370.1422A}
{Altay} G.,  {Colberg} J.~M.,   {Croft} R.~A.~C.,  2006, \mn@doi [\mnras]
  {10.1111/j.1365-2966.2006.10555.x}, \href
  {http://adsabs.harvard.edu/abs/2006MNRAS.370.1422A} {370, 1422}

\bibitem[\protect\citeauthoryear{{Aragon-Calvo} \& {Yang}}{{Aragon-Calvo} \&
  {Yang}}{2014}]{2014MNRAS.440L..46A}
{Aragon-Calvo} M.~A.,  {Yang} L.~F.,  2014, \mn@doi [\mnras]
  {10.1093/mnrasl/slu009}, \href
  {http://adsabs.harvard.edu/abs/2014MNRAS.440L..46A} {440, L46}

\bibitem[\protect\citeauthoryear{{Arag{\'o}n-Calvo}, {van de Weygaert}, {Jones}
   \& {van der Hulst}}{{Arag{\'o}n-Calvo} et~al.}{2007}]{2007ApJ...655L...5A}
{Arag{\'o}n-Calvo} M.~A.,  {van de Weygaert} R.,  {Jones} B. J.~T.,   {van der
  Hulst} J.~M.,  2007, \mn@doi [\apj] {10.1086/511633}, \href
  {https://ui.adsabs.harvard.edu/#abs/2007ApJ...655L...5A} {655, L5}

\bibitem[\protect\citeauthoryear{{Arag{\'o}n-Calvo}, {van de Weygaert}  \&
  {Jones}}{{Arag{\'o}n-Calvo} et~al.}{2010}]{2010MNRAS.408.2163A}
{Arag{\'o}n-Calvo} M.~A.,  {van de Weygaert} R.,   {Jones} B.~J.~T.,  2010,
  \mn@doi [\mnras] {10.1111/j.1365-2966.2010.17263.x}, \href
  {http://adsabs.harvard.edu/abs/2010MNRAS.408.2163A} {408, 2163}

\bibitem[\protect\citeauthoryear{{Blanton}, {Lin}, {Lupton}, {Maley}, {Young},
  {Zehavi}\  \& {Loveday}}{{Blanton} et~al.}{2003}]{Blanton:2003}
{Blanton} M.~R.,  {Lin} H.,  {Lupton} R.~H.,  {Maley} F.~M.,  {Young} N.,
  {Zehavi}\ I.,   {Loveday} J.,  2003, \mn@doi [\aj] {10.1086/344761}, \href
  {http://adsabs.harvard.edu/abs/2003AJ....125.2276B} {125, 2276}

\bibitem[\protect\citeauthoryear{{Blazek}, {MacCrann}, {Troxel}  \&
  {Fang}}{{Blazek} et~al.}{2017}]{2017arXiv170809247B}
{Blazek} J.,  {MacCrann} N.,  {Troxel} M.~A.,   {Fang} X.,  2017, preprint
  (arXiv:1708.09247), \href {http://adsabs.harvard.edu/abs/2017arXiv170809247B}
  {}

\bibitem[\protect\citeauthoryear{{Bolton} et~al.,}{{Bolton}
  et~al.}{2012}]{Bolton:2012}
{Bolton} A.~S.,  et~al., 2012, \mn@doi [\aj] {10.1088/0004-6256/144/5/144},
  \href {http://adsabs.harvard.edu/abs/2012AJ....144..144B} {144, 144}

\bibitem[\protect\citeauthoryear{{Cautun}, {van de Weygaert}  \&
  {Jones}}{{Cautun} et~al.}{2013}]{Cautun2012}
{Cautun} M.,  {van de Weygaert} R.,   {Jones} B.~J.~T.,  2013, \mn@doi [MNRAS]
  {10.1093/mnras/sts416}, \href
  {http://adsabs.harvard.edu/abs/2013MNRAS.429.1286C} {429, 1286}

\bibitem[\protect\citeauthoryear{{Chen}, {Genovese}  \& {Wasserman}}{{Chen}
  et~al.}{2015a}]{2014arXiv1406.5663C}
{Chen} Y.-C.,  {Genovese} C.~R.,   {Wasserman} L.,  2015a, The Annals of
  Statistics, \href {http://adsabs.harvard.edu/abs/2014arXiv1406.5663C} {43,
  1896}

\bibitem[\protect\citeauthoryear{{Chen}, {Ho}, {Freeman}, {Genovese}  \&
  {Wasserman}}{{Chen} et~al.}{2015b}]{2015MNRAS.454.1140C}
{Chen} Y.-C.,  {Ho} S.,  {Freeman} P.~E.,  {Genovese} C.~R.,   {Wasserman} L.,
  2015b, \mn@doi [\mnras] {10.1093/mnras/stv1996}, \href
  {http://adsabs.harvard.edu/abs/2015MNRAS.454.1140C} {454, 1140}

\bibitem[\protect\citeauthoryear{{Chen} et~al.,}{{Chen}
  et~al.}{2015c}]{2015MNRAS.454.3341C}
{Chen} Y.-C.,  et~al., 2015c, \mn@doi [\mnras] {10.1093/mnras/stv2260}, \href
  {http://adsabs.harvard.edu/abs/2015MNRAS.454.3341C} {454, 3341}

\bibitem[\protect\citeauthoryear{{Chen}, {Ho}, {Brinkmann}, {Freeman},
  {Genovese}, {Schneider}  \& {Wasserman}}{{Chen}
  et~al.}{2016}]{2016MNRAS.461.3896C}
{Chen} Y.-C.,  {Ho} S.,  {Brinkmann} J.,  {Freeman} P.~E.,  {Genovese} C.~R.,
  {Schneider} D.~P.,   {Wasserman} L.,  2016, \mn@doi [\mnras]
  {10.1093/mnras/stw1554}, \href
  {http://adsabs.harvard.edu/abs/2016MNRAS.461.3896C} {461, 3896}

\bibitem[\protect\citeauthoryear{{Conroy}, {Gunn}  \& {White}}{{Conroy}
  et~al.}{2009}]{2009ApJ...699..486C}
{Conroy} C.,  {Gunn} J.~E.,   {White} M.,  2009, \mn@doi [\apj]
  {10.1088/0004-637X/699/1/486}, \href
  {http://adsabs.harvard.edu/abs/2009ApJ...699..486C} {699, 486}

\bibitem[\protect\citeauthoryear{{DES Collaboration} et~al.,}{{DES
  Collaboration} et~al.}{2017}]{2017arXiv170801530D}
{DES Collaboration} et~al., 2017, preprint (arXiv:1708.01530), \href
  {http://adsabs.harvard.edu/abs/2017arXiv170801530D} {}

\bibitem[\protect\citeauthoryear{{Dawson} et~al.,}{{Dawson}
  et~al.}{2013}]{Dawson:2013}
{Dawson} K.~S.,  et~al., 2013, \mn@doi [\aj] {10.1088/0004-6256/145/1/10},
  \href {http://adsabs.harvard.edu/abs/2013AJ....145...10D} {145, 10}

\bibitem[\protect\citeauthoryear{{Forero-Romero}, {Hoffman}, {Gottl{\"o}ber},
  {Klypin}  \& {Yepes}}{{Forero-Romero} et~al.}{2009}]{2009MNRAS.396.1815F}
{Forero-Romero} J.~E.,  {Hoffman} Y.,  {Gottl{\"o}ber} S.,  {Klypin} A.,
  {Yepes} G.,  2009, \mn@doi [\mnras] {10.1111/j.1365-2966.2009.14885.x}, \href
  {http://adsabs.harvard.edu/abs/2009MNRAS.396.1815F} {396, 1815}

\bibitem[\protect\citeauthoryear{{Fukugita}, {Ichikawa}, {Gunn}, {Doi},
  {Shimasaku}  \& {Schneider}}{{Fukugita} et~al.}{1996}]{1996AJ....111.1748F}
{Fukugita} M.,  {Ichikawa} T.,  {Gunn} J.~E.,  {Doi} M.,  {Shimasaku} K.,
  {Schneider} D.~P.,  1996, \mn@doi [\aj] {10.1086/117915}, \href
  {http://adsabs.harvard.edu/abs/1996AJ....111.1748F} {111, 1748}

\bibitem[\protect\citeauthoryear{{Ganeshaiah Veena}, {Cautun}, {van de
  Weygaert}, {Tempel}, {Jones}, {Rieder}  \& {Frenk}}{{Ganeshaiah Veena}
  et~al.}{2018}]{2018arXiv180500033G}
{Ganeshaiah Veena} P.,  {Cautun} M.,  {van de Weygaert} R.,  {Tempel} E.,
  {Jones} B. J.~T.,  {Rieder} S.,   {Frenk} C.~S.,  2018, preprint, \href
  {https://ui.adsabs.harvard.edu/#abs/2018arXiv180500033G} {p.
  arXiv:1805.00033} (\mn@eprint {arXiv} {1805.00033})

\bibitem[\protect\citeauthoryear{{Gunn} et~al.,}{{Gunn}
  et~al.}{1998}]{1998AJ....116.3040G}
{Gunn} J.~E.,  et~al., 1998, \mn@doi [\aj] {10.1086/300645}, \href
  {http://adsabs.harvard.edu/abs/1998AJ....116.3040G} {116, 3040}

\bibitem[\protect\citeauthoryear{{Gunn} et~al.,}{{Gunn}
  et~al.}{2006}]{Gunn2006}
{Gunn} J.~E.,  et~al., 2006, \mn@doi [\aj] {10.1086/500975}, \href
  {http://adsabs.harvard.edu/abs/2006AJ....131.2332G} {131, 2332}

\bibitem[\protect\citeauthoryear{{Hahn}, {Porciani}, {Carollo}  \&
  {Dekel}}{{Hahn} et~al.}{2007a}]{2007MNRAS.375..489H}
{Hahn} O.,  {Porciani} C.,  {Carollo} C.~M.,   {Dekel} A.,  2007a, \mn@doi
  [\mnras] {10.1111/j.1365-2966.2006.11318.x}, \href
  {http://adsabs.harvard.edu/abs/2007MNRAS.375..489H} {375, 489}

\bibitem[\protect\citeauthoryear{{Hahn}, {Carollo}, {Porciani}  \&
  {Dekel}}{{Hahn} et~al.}{2007b}]{2007MNRAS.381...41H}
{Hahn} O.,  {Carollo} C.~M.,  {Porciani} C.,   {Dekel} A.,  2007b, \mn@doi
  [\mnras] {10.1111/j.1365-2966.2007.12249.x}, \href
  {http://adsabs.harvard.edu/abs/2007MNRAS.381...41H} {381, 41}

\bibitem[\protect\citeauthoryear{{Hahn}, {Carollo}, {Porciani}  \&
  {Dekel}}{{Hahn} et~al.}{2007c}]{hahn+07}
{Hahn} O.,  {Carollo} C.~M.,  {Porciani} C.,   {Dekel} A.,  2007c, \mn@doi
  [\mnras] {10.1111/j.1365-2966.2007.12249.x}, \href
  {http://adsabs.harvard.edu/abs/2007MNRAS.381...41H} {381, 41}

\bibitem[\protect\citeauthoryear{{Hildebrandt} et~al.,}{{Hildebrandt}
  et~al.}{2017}]{2017MNRAS.465.1454H}
{Hildebrandt} H.,  et~al., 2017, \mn@doi [\mnras] {10.1093/mnras/stw2805},
  \href {http://adsabs.harvard.edu/abs/2017MNRAS.465.1454H} {465, 1454}

\bibitem[\protect\citeauthoryear{{Hirata} \& {Seljak}}{{Hirata} \&
  {Seljak}}{2003}]{Hirata2003}
{Hirata} C.,  {Seljak} U.,  2003, \mnras, \href
  {http://adsabs.harvard.edu/cgi-bin/nph-bib_query?bibcode=2003MNRAS.343..459H&db_key=AST}
  {343, 459}

\bibitem[\protect\citeauthoryear{{Hogg}, {Finkbeiner}, {Schlegel}  \&
  {Gunn}}{{Hogg} et~al.}{2001}]{2001AJ....122.2129H}
{Hogg} D.~W.,  {Finkbeiner} D.~P.,  {Schlegel} D.~J.,   {Gunn} J.~E.,  2001,
  \aj, \href
  {http://adsabs.harvard.edu/cgi-bin/nph-bib_query?bibcode=2001AJ....122.2129H&amp;db_key=AST}
  {122, 2129}

\bibitem[\protect\citeauthoryear{{Ivezi{\'c}} et~al.,}{{Ivezi{\'c}}
  et~al.}{2004}]{2004AN....325..583I}
{Ivezi{\'c}} {\v Z}.,  et~al., 2004, Astronomische Nachrichten, \href
  {http://adsabs.harvard.edu/cgi-bin/nph-bib_query?bibcode=2004AN....325..583I&db_key=AST}
  {325, 583}

\bibitem[\protect\citeauthoryear{{Joachimi} et~al.,}{{Joachimi}
  et~al.}{2015}]{2015SSRv..193....1J}
{Joachimi} B.,  et~al., 2015, \mn@doi [\ssr] {10.1007/s11214-015-0177-4}, \href
  {http://adsabs.harvard.edu/abs/2015SSRv..193....1J} {193, 1}

\bibitem[\protect\citeauthoryear{{Kilbinger}}{{Kilbinger}}{2015}]{2015RPPh...78h6901K}
{Kilbinger} M.,  2015, \mn@doi [Reports on Progress in Physics]
  {10.1088/0034-4885/78/8/086901}, \href
  {http://adsabs.harvard.edu/abs/2015RPPh...78h6901K} {78, 086901}

\bibitem[\protect\citeauthoryear{{Kirk} et~al.,}{{Kirk}
  et~al.}{2015}]{2015SSRv..193..139K}
{Kirk} D.,  et~al., 2015, \mn@doi [\ssr] {10.1007/s11214-015-0213-4}, \href
  {http://adsabs.harvard.edu/abs/2015SSRv..193..139K} {193, 139}

\bibitem[\protect\citeauthoryear{{Komatsu} et~al.,}{{Komatsu}
  et~al.}{2011}]{Komatsu2011}
{Komatsu} E.,  et~al., 2011, \mn@doi [\apjs] {10.1088/0067-0049/192/2/18},
  \href {http://adsabs.harvard.edu/abs/2011ApJS..192...18K} {192, 18}

\bibitem[\protect\citeauthoryear{{Libeskind}, {Hoffman}, {Forero-Romero},
  {Gottl{\"o}ber}, {Knebe}, {Steinmetz}  \& {Klypin}}{{Libeskind}
  et~al.}{2013}]{2013MNRAS.428.2489L}
{Libeskind} N.~I.,  {Hoffman} Y.,  {Forero-Romero} J.,  {Gottl{\"o}ber} S.,
  {Knebe} A.,  {Steinmetz} M.,   {Klypin} A.,  2013, \mn@doi [\mnras]
  {10.1093/mnras/sts216}, \href
  {http://adsabs.harvard.edu/abs/2013MNRAS.428.2489L} {428, 2489}

\bibitem[\protect\citeauthoryear{{Mandelbaum}}{{Mandelbaum}}{2017}]{2017arXiv171003235M}
{Mandelbaum} R.,  2017, preprint, \href
  {http://adsabs.harvard.edu/abs/2017arXiv171003235M} {} (\mn@eprint {arXiv}
  {1710.03235})

\bibitem[\protect\citeauthoryear{Ozertem \& Erdogmus}{Ozertem \&
  Erdogmus}{2011}]{Ozertem2011}
Ozertem U.,  Erdogmus D.,  2011, Journal of Machine Learning Research, 12, 1249

\bibitem[\protect\citeauthoryear{{Padmanabhan} et~al.,}{{Padmanabhan}
  et~al.}{2008}]{2008ApJ...674.1217P}
{Padmanabhan} N.,  et~al., 2008, \mn@doi [\apj] {10.1086/524677}, \href
  {http://adsabs.harvard.edu/abs/2008ApJ...674.1217P} {674, 1217}

\bibitem[\protect\citeauthoryear{{Reid} et~al.,}{{Reid}
  et~al.}{2016}]{2016MNRAS.455.1553R}
{Reid} B.,  et~al., 2016, \mn@doi [\mnras] {10.1093/mnras/stv2382}, \href
  {http://adsabs.harvard.edu/abs/2016MNRAS.455.1553R} {455, 1553}

\bibitem[\protect\citeauthoryear{{Reyes}, {Mandelbaum}, {Gunn}, {Nakajima},
  {Seljak}  \& {Hirata}}{{Reyes} et~al.}{2012}]{Reyes2012}
{Reyes} R.,  {Mandelbaum} R.,  {Gunn} J.~E.,  {Nakajima} R.,  {Seljak} U.,
  {Hirata} C.~M.,  2012, \mn@doi [\mnras] {10.1111/j.1365-2966.2012.21472.x},
  \href {http://adsabs.harvard.edu/abs/2012MNRAS.425.2610R} {425, 2610}

\bibitem[\protect\citeauthoryear{{Rozo} \& {Rykoff}}{{Rozo} \&
  {Rykoff}}{2014}]{2014ApJ...783...80R}
{Rozo} E.,  {Rykoff} E.~S.,  2014, \mn@doi [\apj] {10.1088/0004-637X/783/2/80},
  \href {http://adsabs.harvard.edu/abs/2014ApJ...783...80R} {783, 80}

\bibitem[\protect\citeauthoryear{{Rozo}, {Rykoff}, {Bartlett}  \&
  {Melin}}{{Rozo} et~al.}{2015}]{2015MNRAS.450..592R}
{Rozo} E.,  {Rykoff} E.~S.,  {Bartlett} J.~G.,   {Melin} J.-B.,  2015, \mn@doi
  [\mnras] {10.1093/mnras/stv605}, \href
  {http://adsabs.harvard.edu/abs/2015MNRAS.450..592R} {450, 592}

\bibitem[\protect\citeauthoryear{{Rykoff} et~al.,}{{Rykoff}
  et~al.}{2014}]{2014ApJ...785..104R}
{Rykoff} E.~S.,  et~al., 2014, \mn@doi [\apj] {10.1088/0004-637X/785/2/104},
  \href {http://adsabs.harvard.edu/abs/2014ApJ...785..104R} {785, 104}

\bibitem[\protect\citeauthoryear{{Schneider} \& {Bridle}}{{Schneider} \&
  {Bridle}}{2010}]{2010MNRAS.402.2127S}
{Schneider} M.~D.,  {Bridle} S.,  2010, \mn@doi [\mnras]
  {10.1111/j.1365-2966.2009.15956.x}, \href
  {http://adsabs.harvard.edu/abs/2010MNRAS.402.2127S} {402, 2127}

\bibitem[\protect\citeauthoryear{{Singh}, {Mandelbaum}  \& {More}}{{Singh}
  et~al.}{2015}]{Singh2015}
{Singh} S.,  {Mandelbaum} R.,   {More} S.,  2015, \mn@doi [\mnras]
  {10.1093/mnras/stv778}, \href
  {http://adsabs.harvard.edu/abs/2015MNRAS.450.2195S} {450, 2195}

\bibitem[\protect\citeauthoryear{{Smee} et~al.,}{{Smee}
  et~al.}{2013}]{Smee:2013}
{Smee} S.~A.,  et~al., 2013, \mn@doi [\aj] {10.1088/0004-6256/146/2/32}, \href
  {http://adsabs.harvard.edu/abs/2013AJ....146...32S} {146, 32}

\bibitem[\protect\citeauthoryear{{Smith} et~al.,}{{Smith}
  et~al.}{2002}]{2002AJ....123.2121S}
{Smith} J.~A.,  et~al., 2002, \mn@doi [\aj] {10.1086/339311}, \href
  {http://adsabs.harvard.edu/abs/2002AJ....123.2121S} {123, 2121}

\bibitem[\protect\citeauthoryear{{Sousbie}, {Pichon}, {Colombi}, {Novikov}  \&
  {Pogosyan}}{{Sousbie} et~al.}{2008}]{2008MNRAS.383.1655S}
{Sousbie} T.,  {Pichon} C.,  {Colombi} S.,  {Novikov} D.,   {Pogosyan} D.,
  2008, \mn@doi [\mnras] {10.1111/j.1365-2966.2007.12685.x}, \href
  {http://adsabs.harvard.edu/abs/2008MNRAS.383.1655S} {383, 1655}

\bibitem[\protect\citeauthoryear{{Tempel} \& {Libeskind}}{{Tempel} \&
  {Libeskind}}{2013}]{2013ApJ...775L..42T}
{Tempel} E.,  {Libeskind} N.~I.,  2013, \mn@doi [\apjl]
  {10.1088/2041-8205/775/2/L42}, \href
  {http://adsabs.harvard.edu/abs/2013ApJ...775L..42T} {775, L42}

\bibitem[\protect\citeauthoryear{{Tempel}, {Stoica}  \& {Saar}}{{Tempel}
  et~al.}{2013}]{2013MNRAS.428.1827T}
{Tempel} E.,  {Stoica} R.~S.,   {Saar} E.,  2013, \mn@doi [\mnras]
  {10.1093/mnras/sts162}, \href
  {http://adsabs.harvard.edu/abs/2013MNRAS.428.1827T} {428, 1827}

\bibitem[\protect\citeauthoryear{{Tempel}, {Guo}, {Kipper}  \&
  {Libeskind}}{{Tempel} et~al.}{2015}]{2015MNRAS.450.2727T}
{Tempel} E.,  {Guo} Q.,  {Kipper} R.,   {Libeskind} N.~I.,  2015, \mn@doi
  [\mnras] {10.1093/mnras/stv919}, \href
  {http://adsabs.harvard.edu/abs/2015MNRAS.450.2727T} {450, 2727}

\bibitem[\protect\citeauthoryear{Wasserman}{Wasserman}{2006}]{wasserman2006all}
Wasserman L.,  2006, All of Nonparametric Statistics.
Springer-Verlag New York, Inc.

\bibitem[\protect\citeauthoryear{{York} et~al.,}{{York}
  et~al.}{2000}]{2000AJ....120.1579Y}
{York} D.~G.,  et~al., 2000, \mn@doi [\aj] {10.1086/301513}, \href
  {http://adsabs.harvard.edu/abs/2000AJ....120.1579Y} {120, 1579}

\bibitem[\protect\citeauthoryear{{Zhang}, {Yang}, {Faltenbacher}, {Springel},
  {Lin}  \& {Wang}}{{Zhang} et~al.}{2009}]{2009ApJ...706..747Z}
{Zhang} Y.,  {Yang} X.,  {Faltenbacher} A.,  {Springel} V.,  {Lin} W.,   {Wang}
  H.,  2009, \mn@doi [\apj] {10.1088/0004-637X/706/1/747}, \href
  {http://adsabs.harvard.edu/abs/2009ApJ...706..747Z} {706, 747}

\bibitem[\protect\citeauthoryear{{Zhang}, {Yang}, {Wang}, {Wang}, {Mo}  \& {van
  den Bosch}}{{Zhang} et~al.}{2013}]{2013ApJ...779..160Z}
{Zhang} Y.,  {Yang} X.,  {Wang} H.,  {Wang} L.,  {Mo} H.~J.,   {van den Bosch}
  F.~C.,  2013, \mn@doi [\apj] {10.1088/0004-637X/779/2/160}, \href
  {http://adsabs.harvard.edu/abs/2013ApJ...779..160Z} {779, 160}

\bibitem[\protect\citeauthoryear{{Zhang}, {Yang}, {Wang}, {Wang}, {Luo}, {Mo}
  \& {van den Bosch}}{{Zhang} et~al.}{2015}]{2015ApJ...798...17Z}
{Zhang} Y.,  {Yang} X.,  {Wang} H.,  {Wang} L.,  {Luo} W.,  {Mo} H.~J.,   {van
  den Bosch} F.~C.,  2015, \mn@doi [\apj] {10.1088/0004-637X/798/1/17}, \href
  {http://adsabs.harvard.edu/abs/2015ApJ...798...17Z} {798, 17}

\makeatother
\end{thebibliography}

\appendix

\section{Effect from filament uncertainty}	\label{sec::smooth}

\begin{figure}
\center
\includegraphics[height=2.4in]{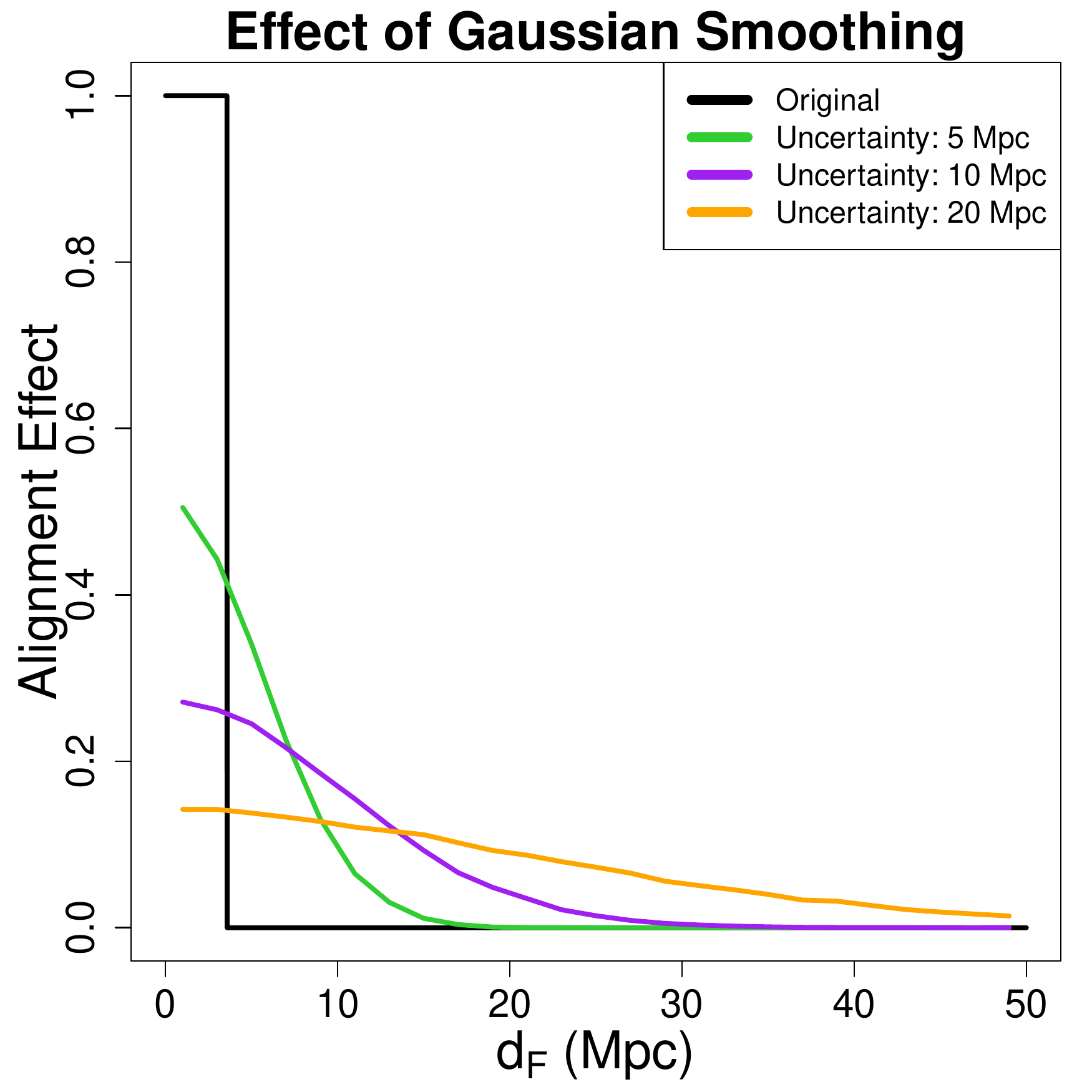}
\caption{
A simple model to analyze the effect of filament uncertainty on the observed galaxy-filament
alignment effect. 
The black curve describes a simple model for the alignment effect: 
within a distance $3.5$ Mpc, the effect has a magnitude $1$ and $0$ otherwise.
We then smooth this effect by a different amount of Gaussian uncertainty ($5$, $10$, and $20$ Mpc)
and display how the alignment effect may be extended to a longer distance.
The uncertainty of filaments in the Cosmic Web Reconstruction is around $10-20$ Mpc in the LOWZ sample \citep{2016MNRAS.461.3896C}.
Based on this simple model, the alignment effect may last to $30-50$ Mpc, which is consistent with what we observed in 
\autoref{fig::align}.
}
\label{fig::smooth}
\end{figure}

To investigate the effect from filament uncertainty, we consider a very simple model where
the alignment effect is 
$$
\begin{cases}
1, \quad \mbox{if $d_F\leq 3.5$ Mpc},\\
0, \quad \mbox{if $d_F> 3.5$ Mpc}.
\end{cases}
$$
The threshold $3.5$ Mpc is from the study in simulations in \cite{2015MNRAS.454.3341C}.
To model the impact of uncertainty in the filament position,
we assume that our observed distance to filament 
{{
$$
d_{F, {\sf obs}} = |d_{F} + \epsilon|, 
$$
}
where $d_F$ is the true distance to filament
and $\epsilon $ is a random number follows from a Normal distribution
with variance $\sigma^2.$
We consider $\sigma=5$, $10$, and $20$ Mpc
and we display the result on the alignment signal in \autoref{fig::smooth}.
The uncertainty of filaments in the LOWZ sample is about $10-20$ Mpc according to
\cite{2016MNRAS.461.3896C}, model predictions with $\sigma=10$ and $20$ Mpc provide a rough idea for
how 
the Gaussian uncertainty  distance to the filament affects the range of effect.
According to
\autoref{fig::smooth}
the uncertainty of filaments may lead to an alignment effect with a range up to $50$ Mpc,
which is close to what we observed in \autoref{fig::align}. 
{{Moreover, the magnitude reflects how the alignment effect was damped by the uncertainty.
When the uncertainty is around $10-20$ Mpc, the alignment
effect is reduced to $1/4-1/6$ of the original alignment effect.}

\section{Checking for the systematics}	\label{sec::system}

In this section, we investigate possible systematics using two approaches.
The first one is to study the systematics from the distribution of filaments' orientations ($\phi_F$).
The other one is to study the alignment of galaxies when they are far away from both filaments and galaxies.

%We examine the systematics from the distribution of filaments' orientations ($\phi_F$). 
Ideally, the distribution of filaments' orientations should be close to a uniform distribution.
However, as is shown in the top-left panel of \autoref{fig::check2}, 
the distribution has bumps at $\pm 90\deg$ ($\phi_F = \pm 90\deg$, the direction along RA direction), 
indicating that
we have more filaments along the RA direction than the dec direction. 
This is probably due to 
the artifact of the survey geometry, where 
%This is probably due to the limitation of our filament finder. 
%We detect filaments using ridgelines of the kernel density estimator. 
%The statistical bias of the kernel density estimator tends to
%be higher around the boundary of the data. 
%In the SDSS data, 
the ranges of RA is much wider than the range of dec,
making the boundary along RA direction larger than along the dec direction.
Thus, our filament finders will tend to detect more filaments along the long boundary (RA direction)
than the short boundary (dec direction). 
{To further examine the boundary bias, we consider galaxies and filaments within the region $150\degr<RA<200\degr$ and 
$10\degr<dec<30\degr$. This region is chosen because it is in the interior of the LOWZ sample area.
We compute the orientations of filaments in the top-right panel.
The two bumps around $\pm 90\deg$ get damped and the distribution
is more randomly fluctuating, which suggest that the boundary bias may be 
the main driving force for the two bumps. 
Note that the fluctuation in this panel may be due to the cosmic variance since we 
are using a small region in the sky.
}

%so the boundary bias will have a stronger impact on the RA direction. 
%Note that the $\phi_F$ is within $[-90,90]$ degree because when
%we compute the orientation of filament, we force it to be within this range. 
%In the middle panel of \autoref{fig::check2}, we display the distribution of $\phi_{\sf Major}$,
%which 

Although there are artifacts in filaments' orientations,
the distribution of $\phi_{\sf Major}$ is very uniformly distributed (bottom-left panel of \autoref{fig::check2}). 
To investigate if the artifacts of filaments' orientation will cause systematics in
galaxy-filament alignments,
we randomly shuffle the redshift slices of filaments and compute the alignment signal.
Namely, galaxies in redshift slice A may be compared to filaments from redshift slice B. 
If there is no systematics, we will not observe any alignment signal. 
In the bottom-right panel of \autoref{fig::check2}, 
we see that the result is consistently with random orientations, which suggests
{
that there is no strong evidence of systematics from different redshift slices.
}
%that there is no systematics in the alignment signals.

{
To examine if there are systematics within the same redshift slice,
we randomly pair a galaxy to a filament within the same slice, regardless of their distance.
For each pair of galaxy-filament, we compute the angular difference
between the galaxy's major axis and the filament's orientation. 
The result is given in the left panel of Figure~\ref{fig::check3}.
Clearly, the result is consistent with a flat line, suggesting that
there is no evidence about the systematics within the same redshift slice.
As a reference, we attach the result when we pair a galaxy to the nearest filament in the 
right panel (this is our alignment signal).
We observed a  significant increase of probability in the aligned cases (low value of $|\phi_F-\phi_{\sf Major}|$),
which suggests the presence of galaxy-filament alignments. 
}

{
Finally, to examine possible systematics caused by coordinate transformations,
we separate the data into two parts according the declination: 
one sample with a high declination ($>24\degr$)
and the other one with a low declination ($<24\degr$). 
Note that the threshold $24$ degrees is chosen based on the median declination of the entire LOWZ sample.
For each sample, we apply the same technique to detect the alignment signal and
the result is given in Figure~\ref{fig::check4}.
In both cases, we observe a significant alignment when the distance to filament is small
and the alignment disappears when the distance to filament is large, which is consistent with
the observations in Figure~\ref{fig::align}.
And the trends in both panels are very consistent. 
The fact that there is no clear difference between the low declination and high declination samples
suggest that no systematics from coordinate transformations.
}

%we randomly pair a galaxy and a filament and compute the difference between the
%major axis and filament's orientation in the right panel of \autoref{fig::check2}.
%If there are systematics, the resulting distribution should be non-uniform. 
%We obtain a uniform distribution, which indicates that
%the artifacts of filaments' preferred orientations does not cause systematics in
%the alignment signal.
%Thus, though filaments' orientations have a preferred orientation, 
%this systematics does not bias the alignment signal.
%

\begin{figure*}
\center
\includegraphics[height=2in]{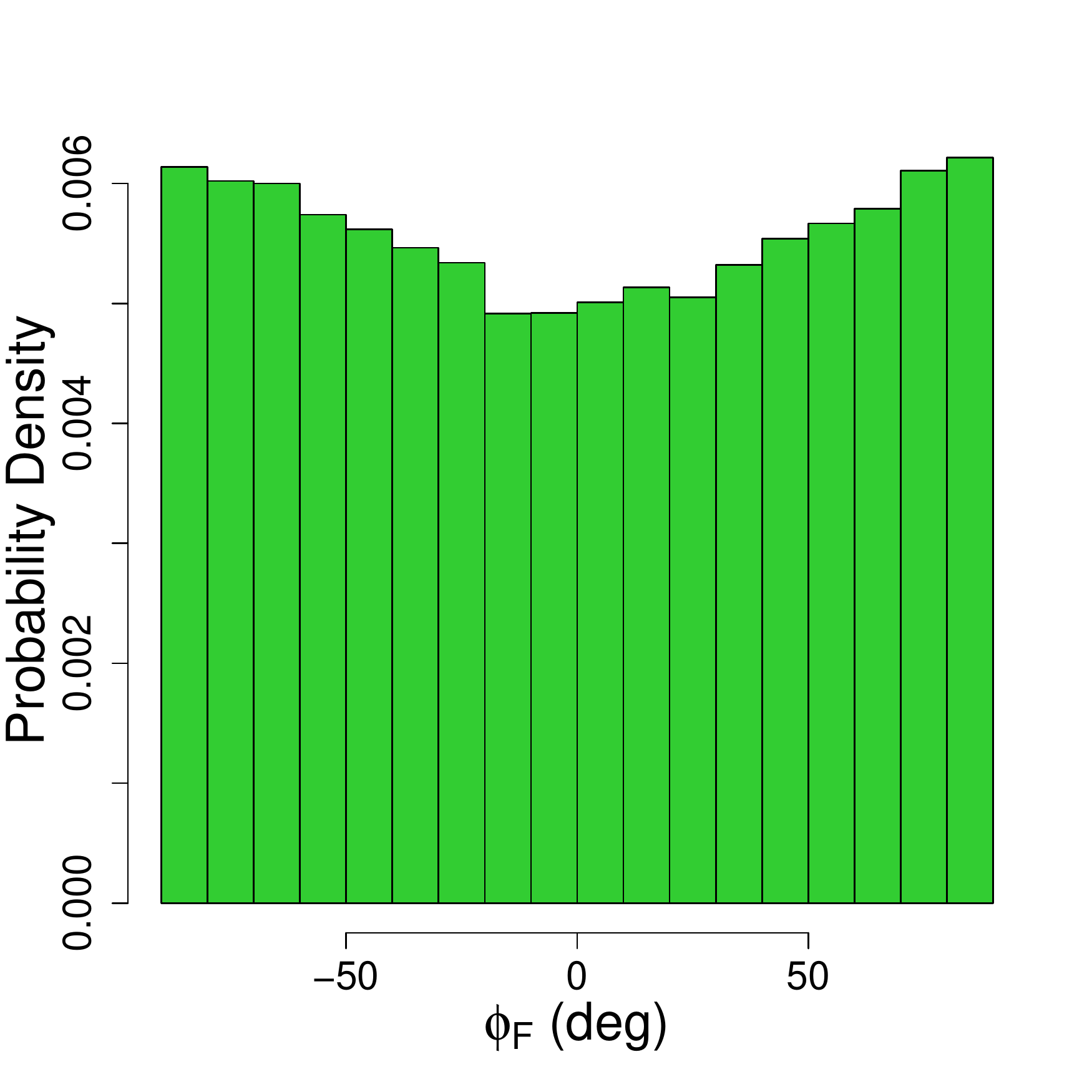}
\includegraphics[height=2in]{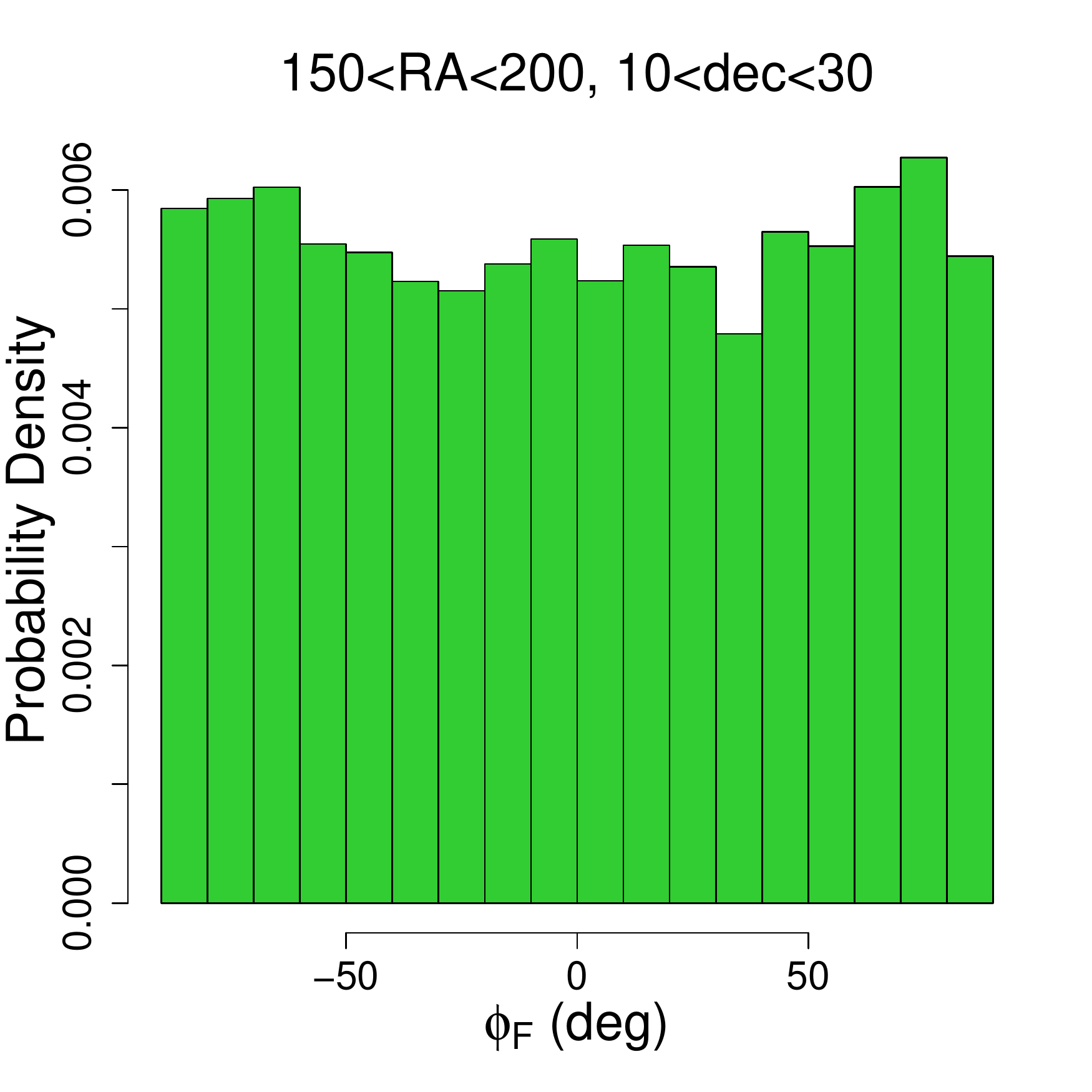}\\
\includegraphics[height=2in]{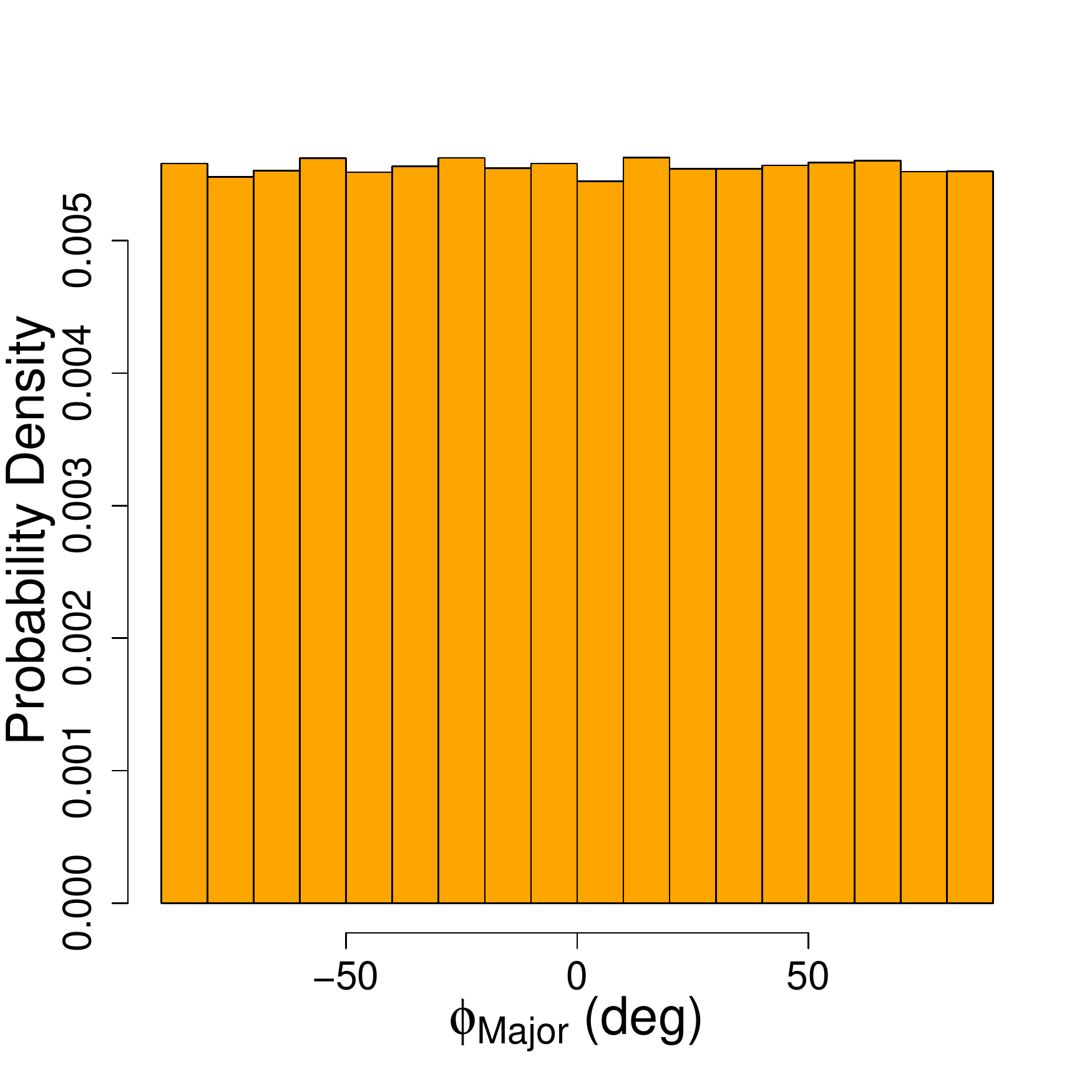}
\includegraphics[height=2in]{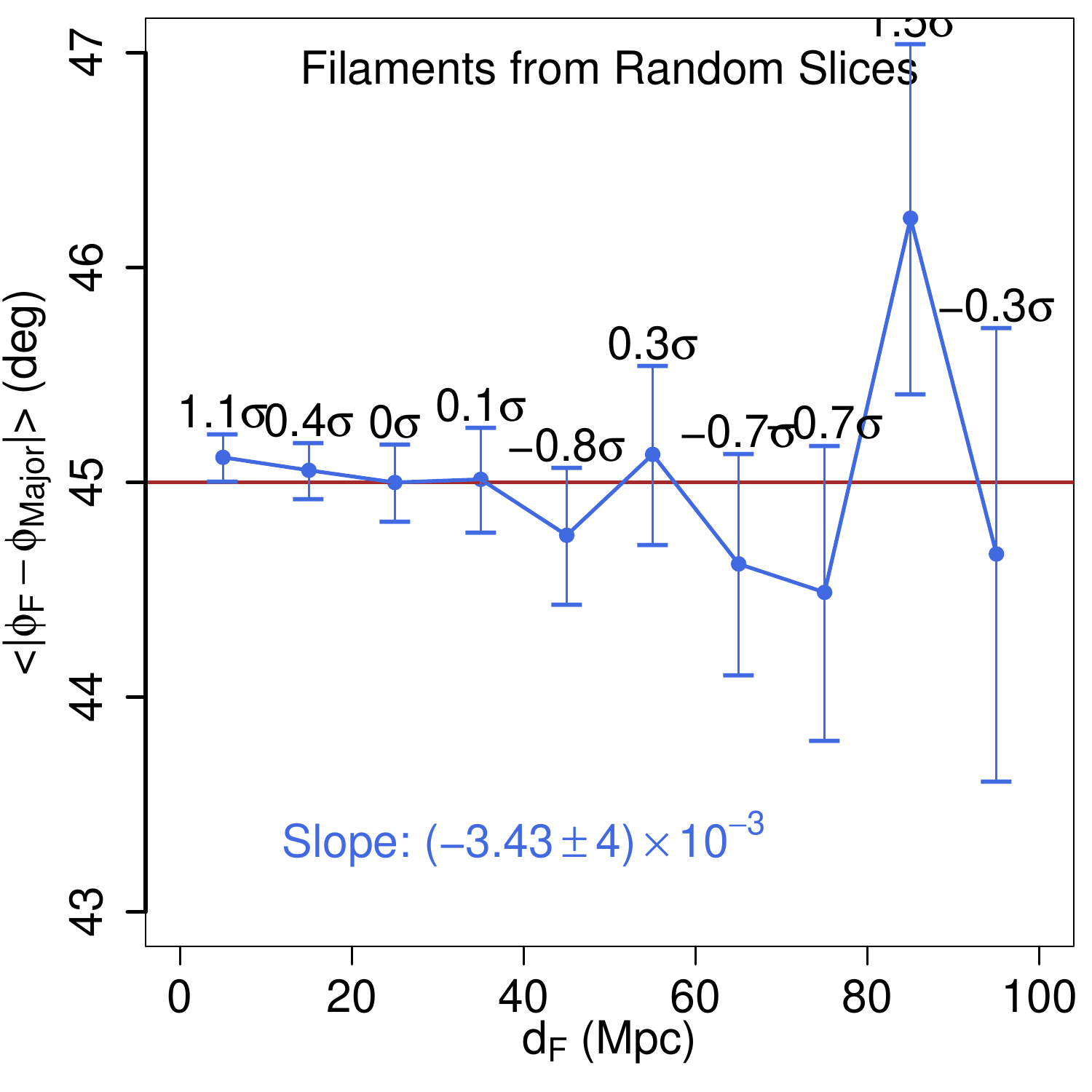}
\caption{
We examine the distributions of filaments' orientations ($\phi_F$) and
the major axes of galaxies ($\phi_{\sf Major}$). 
In the top-left panel, we see that the orientation of filaments are not very uniformly distributed
but instead, there are more filaments with an orientation parallel to the RA direction (close to $\pm 90$ degrees).
This may be caused by the boundary bias of the filament finder -- ridgelines are constructed
by applying the kernel density estimator, which is known to have a higher statistical bias around
the boundary of the data. 
{In the top-right panel, we attempt to 
study examine the boundary bias by focusing the region inside $150<RA<200$, $10<dec<30$
and computing the filaments' orientation. 
This regions is completely inside the survey area of the LOWZ sample
so the boundary bias will be limited. 
Although there are some fluctuations in this panel, there is no clear pattern as in the top-left panel. 
The fluctuation may be caused by cosmic variance. }
In the bottom-left panel, we display the distribution of major axes of galaxies, which follows a uniform distribution.
In the bottom-right panel, we randomly shuffle the redshift slices of filaments and calculate the alignment signal.
We observe a flat line, indicating that there is no systematics in our analysis. 
%In the right panel, we randomly pair a galaxy and a filament and compare their orientation to investigate the 
%systematics.
%The angular distribution is very uniform as well, indicating that the systematics from filaments' orientation
%may not have a strong impact on the alignment signal.
}
\label{fig::check2}
\end{figure*}

\begin{figure*}
\center
\includegraphics[height=2in]{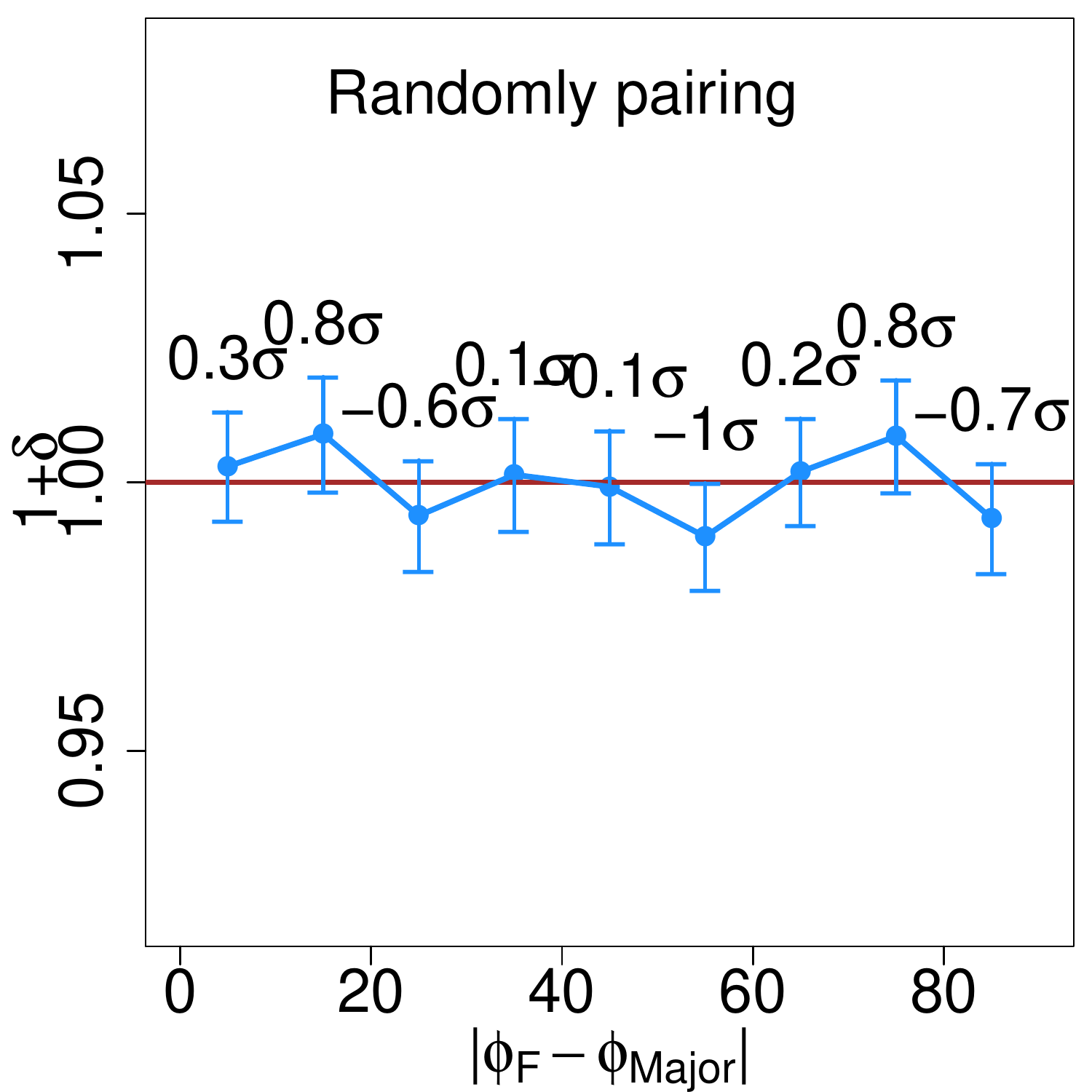}
\includegraphics[height=2in]{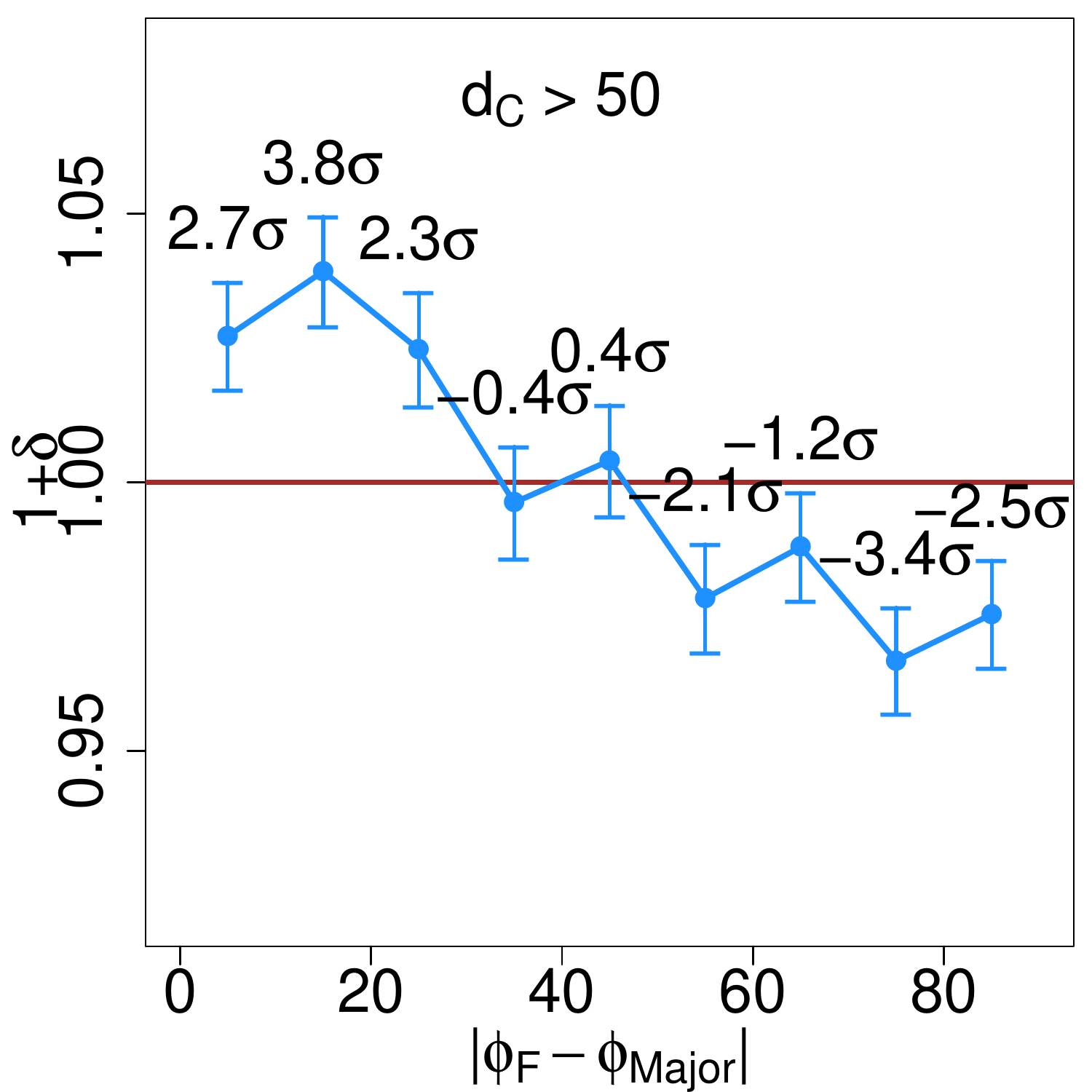}
\caption{
Examining the within-slice systematics using excess probability distribution ($\delta = 1-\frac{p_{\sf obs}}{p_{\sf rand}}$). 
{\bf Right panel:}
To examine the systematics, we randomly pair a galaxy to a filament within the same redshift slice
and compute the angular difference between the major axis of the galaxy and the orientation of the filament.
Here, the excess probability distribution is consistent with a flat line, suggesting that
there is no strong evidence of systematics. 
{\bf Left panel:} This is a reference to the right panel. 
This panel shows the results when we pair a galaxy to the nearest filament
(with a constraint that galaxies have to be at least $50$ Mpc away from clusters to eliminate the effect
from clusters).
%in our observed data for galaxies that are at least $50$ Mpc away from the nearest clusters 
%(we do not impose any constraints on the distance to nearest filament).
We observe an increase in probability in aligned cases (low value of $|\phi_F - \phi_{\sf Major}|$),
suggesting the presence of galaxy-filament alignments.}
\label{fig::check3}
\end{figure*}

\begin{figure*}
\center
\includegraphics[height=2in]{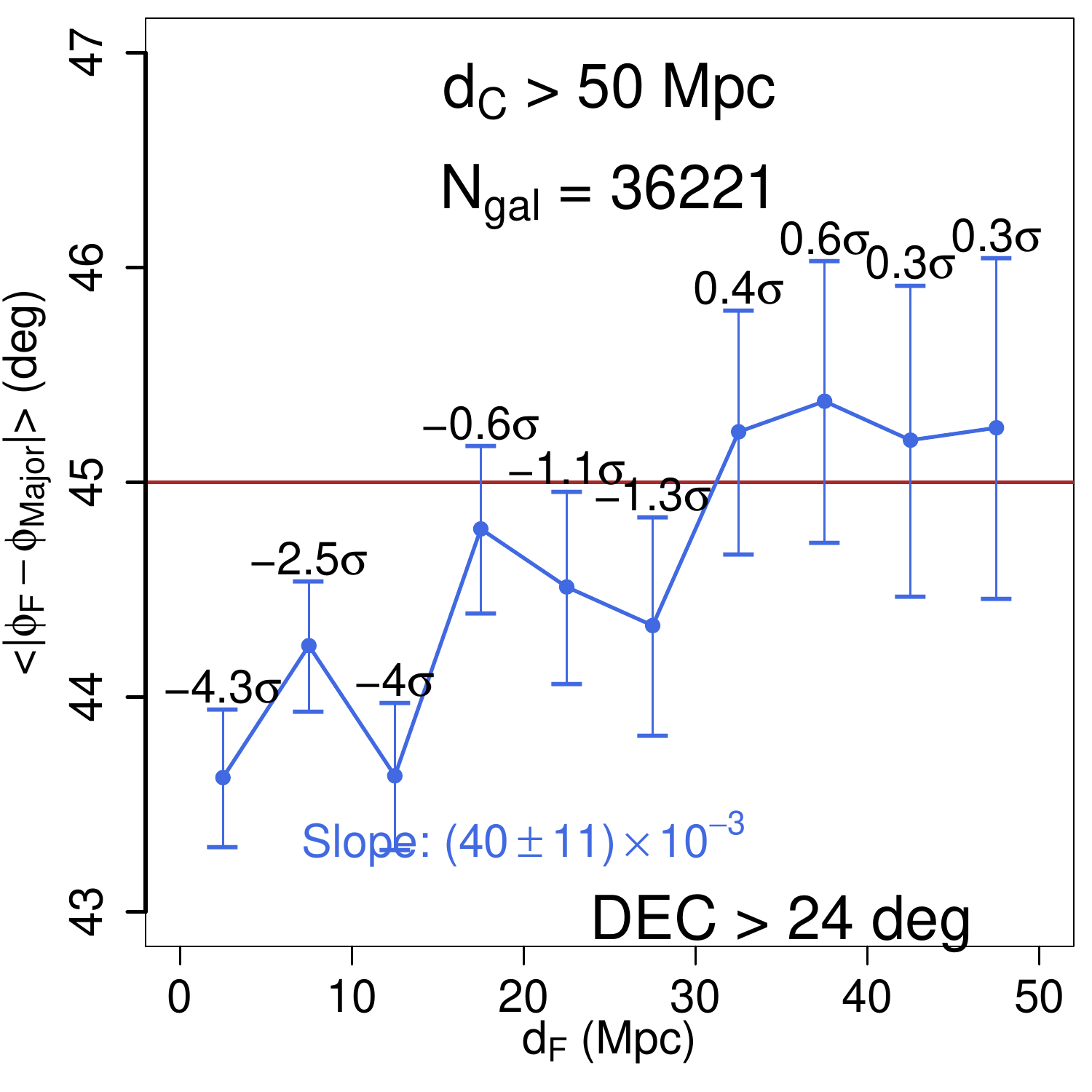}
\includegraphics[height=2in]{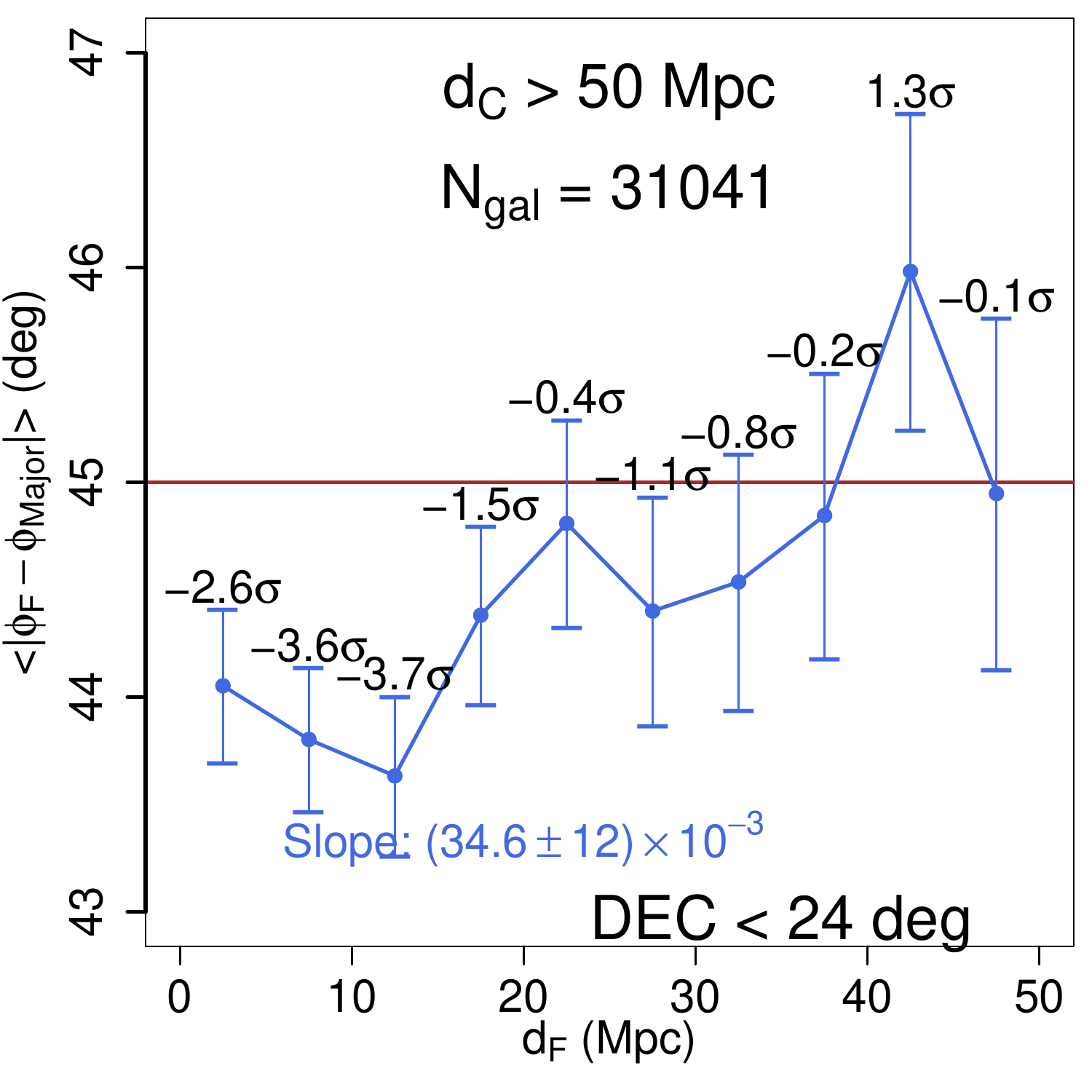}
\caption{
Examining the systematics from coordinate transformations by
splitting the data according to the declination. 
We separate the galaxy sample by the declination into two samples:
a high declination sample and a low declination sample.
We then compute the alignment signal as Figure~\ref{fig::align}. 
We observe a consistent pattern in both panels and the trend is very quantitatively 
similar to the one described in Figure~\ref{fig::align}. 
}
\label{fig::check4}
\end{figure*}

\section{Checking for the anti-alignment}	\label{sec::anti}

In \autoref{fig::NP1},
we notice that there seems to be an anti-alignment effect when $d_F$ is large
(however, it is not statistically significant). 
To investigate if such an anti-alignment is realistic, we analyze the alignments of galaxies toward filaments
for those galaxies that are far away from both filaments and clusters. 

%To further investigate if such an effect exists or if there are systematics inside our analysis,
%we examine the alignment distribution for those galaxies
%that are far away from both clusters and filaments. 
In particular, we consider galaxies with $d_C>50$ Mpc
and $d_F> 50, 100$ Mpc
and investigate the distribution of $\phi_{\sf Major}-\phi_{F}$.
The result is given in
in \autoref{fig::check}.
The distribution is very similar to a uniform distribution in both cases
and the KS test gives p-values $0.07$ (when we threshold $d_F>50$ Mpc) and $0.86$ 
(when we threshold $d_F>100$ Mpc). 
Thus, we do not observe a significant effect in our sample
so we cannot conclude if there is indeed an anti-alignment effect when galaxies are far away from filaments. 
%and
%the anti-alignment at large $d_F$ might be caused by random fluctuations. 

%although $\phi_F$ has a preferential orientation, $\phi_{\sf Major}$
%does not. Thus, when we consider the difference between the two axes, 
%the systematics from $\phi_F$ will be canceled out by the uniform distribution of $\phi_{\sf Major}$.

\begin{figure*}
\center
\includegraphics[height=2.4in]{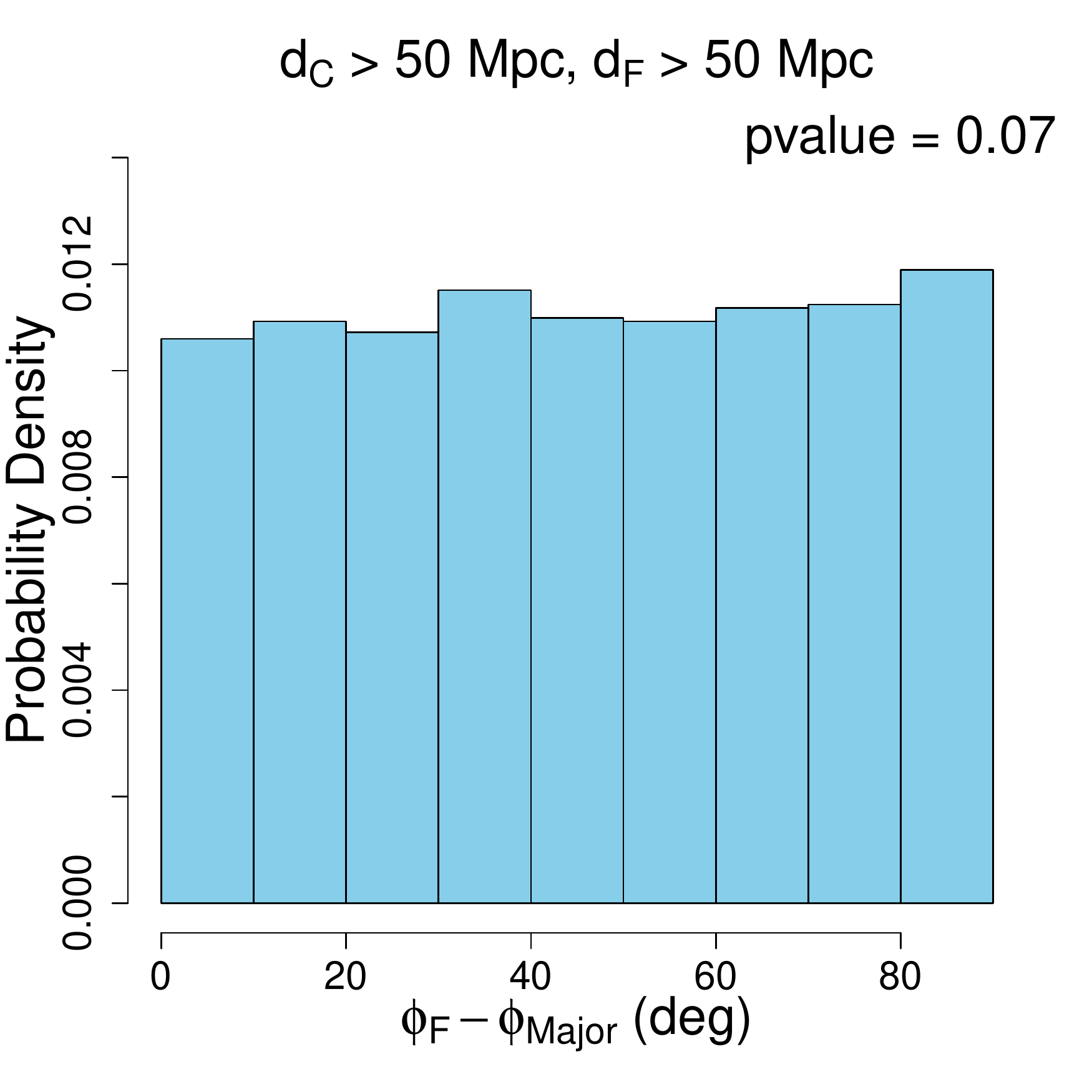}
\includegraphics[height=2.4in]{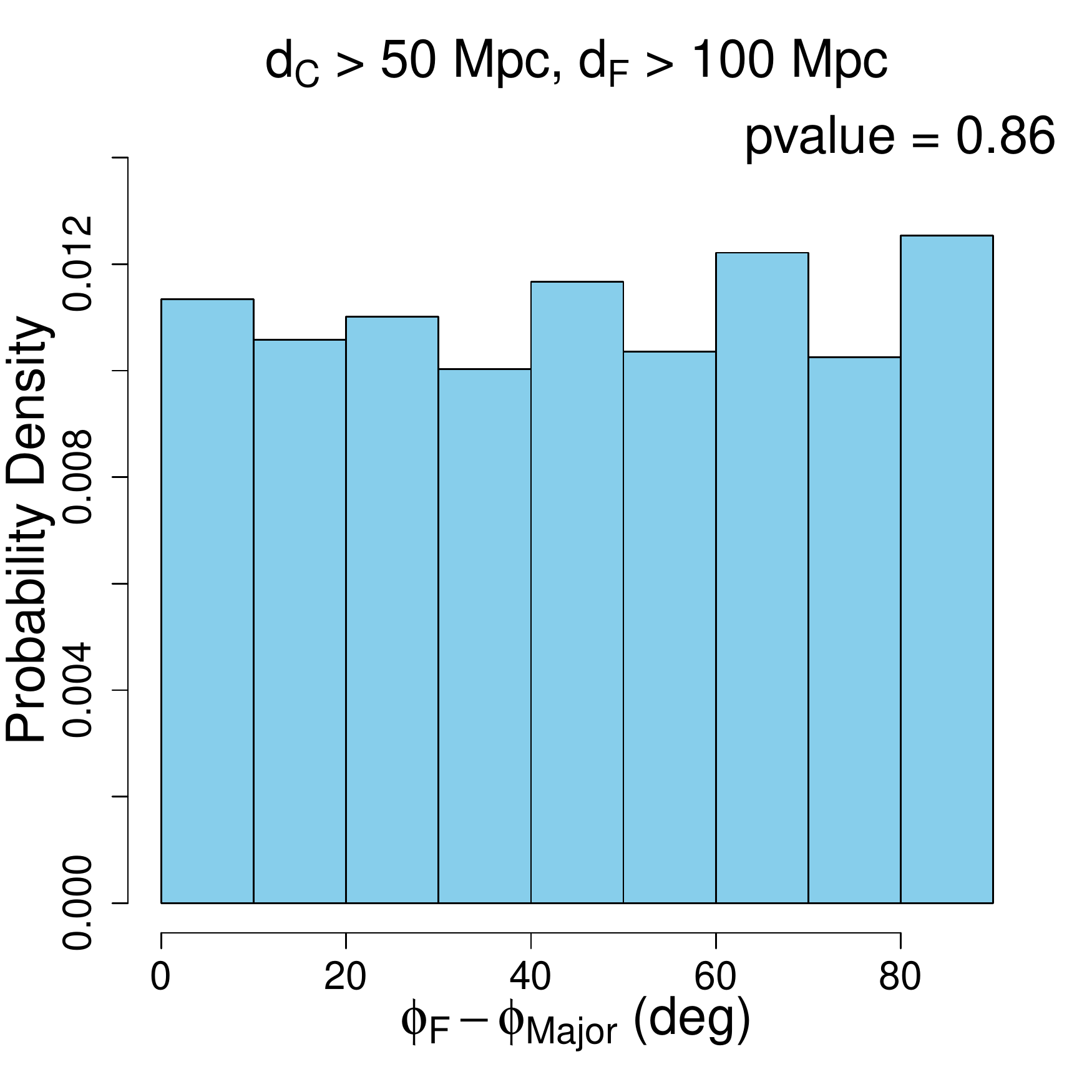}
\caption{
We check the systematics by considering galaxies that are far away from both clusters and filaments.
In the top panel, we consider galaxies with $d_C>50$ Mpc and $d_F>50$ Mpc
and in the bottom panel, we investigate those with $d_C>50$ Mpc and $d_F>100$ Mpc. 
We apply the KS-test to check if the distribution is significantly different from $0$. 
The p-values are $0.07$ and $0.86$, both are not significant to the standard significance level $0.05$.
}
\label{fig::check}
\end{figure*}

\section{P-values from other tests}	\label{sec::other}

In addition to the $\chi^2$ test, we also employ
the KS test and AD test 
to investigate if galaxy's properties have a significant impact on the alignment signal. 
We focus on the comparison in \autoref{sec::property} and \autoref{fig::property}
and
re-create \autoref{tab::test1} with the other two tests.
The p-values are given in \autoref{tab::test_many}. 
Note that for the ease of comparison, we also include the results from \autoref{tab::test1} ($\chi^2$ test).

In every test, there are six scenarios where
a significant difference is observed:
separating galaxies by the most extreme $25\%$ or $15\%$ brightness, 
the most extreme $15\%$ stellar mass, 
and the most extreme $25\%, 15\%$, or $10\%$ age. 
In particular, when galaxies are separated by their brightness
or age,
we see very significant differences in some cases (for instance, the $25\%$ extreme brightness 
or the $15\%$ extreme age comparison) across every test.

\begin{table}
\centering
\begin{tabular}{crrr}
  \hline
(KS test) & 25\% & 15\% & 10\% \\ 
  \hline
%Brightness & 27.54 & 21.89  & 14.93  \\ 
%Stellar Mass & 8.7  & 14.88  & 9.65  \\ 
%  Age & 16.42  & 25.81  & 13.83  \\ 
Brightness & $\mathbf{3.6\times 10^{-3}}$ & $\mathbf{2.6\times 10^{-2}}$  & ${1.3\times 10^{-1}}$  \\ 
Stellar Mass & $1.1\times 10^{-1}$  & $\mathbf{2.2\times10^{-2}}$  & $7.7\times10^{-2}$  \\ 
Age & $\mathbf{4.4\times10^{-2}}$  & $\mathbf{2.8\times10^{-3}}$  & $\mathbf{1.7\times10^{-3}}$  \\ 
   \hline
\end{tabular}
\begin{tabular}{crrr}
  \hline
(AD test) & 25\% & 15\% & 10\% \\ 
  \hline
%Brightness & 27.54 & 21.89  & 14.93  \\ 
%Stellar Mass & 8.7  & 14.88  & 9.65  \\ 
%  Age & 16.42  & 25.81  & 13.83  \\ 
Brightness & $\mathbf{4.0\times 10^{-4}}$ & $\mathbf{6.3\times 10^{-3}}$  & ${7.1\times 10^{-2}}$  \\ 
Stellar Mass & $\mathbf{4.1\times 10^{-2}}$  & $\mathbf{4.1\times10^{-3}}$  & $5.6\times10^{-2}$  \\ 
Age & $\mathbf{2.5\times10^{-3}}$  & $\mathbf{4.3\times10^{-3}}$  & $\mathbf{4.7\times10^{-3}}$  \\ 
   \hline
\end{tabular}
\begin{tabular}{crrr}
  \hline
($\chi^2$ test) & 25\% & 15\% & 10\% \\ 
  \hline
%Brightness & 27.54 & 21.89  & 14.93  \\ 
%Stellar Mass & 8.7  & 14.88  & 9.65  \\ 
%  Age & 16.42  & 25.81  & 13.83  \\ 
%Brightness & $\mathbf{1.1\times 10^{-3}}$ & $\mathbf{3.6\times 10^{-3}}$  & $\mathbf{2.9\times 10^{-2}}$  \\ 
%Stellar Mass & $6.1\times 10^{-2}$  & $\mathbf{1.5\times10^{-2}}$  & $8.5\times10^{-2}$  \\ 
%  Age & $\mathbf{6.9\times10^{-3}}$  & $\mathbf{1.6\times10^{-3}}$  & $\mathbf{2.9\times10^{-2}}$  \\ 
Brightness & $\mathbf{7.1\times 10^{-4}}$ & $\mathbf{4.8\times 10^{-3}}$  & $\mathbf{2.9\times 10^{-2}}$  \\ 
Stellar Mass & $7.5\times 10^{-2}$  & $\mathbf{1.1\times10^{-2}}$  & $5.0\times10^{-2}$  \\ 
  Age & $\mathbf{3.0\times10^{-2}}$  & $\mathbf{3.3\times10^{-3}}$  & ${6.8\times10^{-2}}$  \\ 
   \hline
\end{tabular}
\caption{
Test for the significance between the two most extreme types of galaxies by different properties
with different testing approaches. 
In the top table, we use the KS-test. In the middle table, we use the AD test.
In the bottom table, we use the $\chi^2$ test, which is the same as Table~\ref{tab::test1}.
}
\label{tab::test_many}

\end{table}

\bsp

\label{lastpage}

\end{document}